\documentclass{aa}
\usepackage{graphicx}
\usepackage{xcolor}
\usepackage{listings}
\usepackage{subfig}
\usepackage[bookmarks=false]{hyperref}

\newcommand{\serval}{{\tt serval}}
\newcommand{\molecfit}{{\tt molecfit}}
\newcommand{\raccoon}{{\tt raccoon}}

\hypersetup{
     colorlinks = true,
     linkcolor = blue,
     anchorcolor = blue,
     citecolor = blue,
     filecolor = blue,
     urlcolor = blue
     }
\definecolor{dkgreen}{rgb}{0,0.6,0}
\definecolor{gray}{rgb}{0.5,0.5,0.5}
\definecolor{mauve}{rgb}{0.58,0,0.82}

\usepackage{txfonts}
\usepackage{cleveref}
\usepackage[LGRgreek]{mathastext}

\begin{document} 

\title{The widest broadband transmission spectrum (0.38--1.71\,$\mu$m) \\ of HD~189733b from ground-based chromatic \\ Rossiter-McLaughlin observations}
   
\author{M.~Oshagh\inst{\ref{iiac},\ref{iull},\ref{iiag}}, 
F.\,F. Bauer\inst{\ref{iiaa}}, 
M.~Lafarga\inst{\ref{iice},\ref{iieec}}, 
K.~Molaverdikhani\inst{\ref{impia},\ref{ilza}}, 
P.\,J.~Amado\inst{\ref{iiaa}}, 
L.~Nortmann\inst{\ref{iiag}}, 
A.~Reiners\inst{\ref{iiag}}, 
A.~Guzm\'an-Mesa\inst{\ref{iiag},\ref{icsh}}, 
E.~Pall\'e\inst{\ref{iiac},\ref{iull}},
E.~Nagel\inst{\ref{ihamb},\ref{itaut}},
J.\,A.~Caballero\inst{\ref{iesac}},
N.~Casasayas-Barris\inst{\ref{iiac},\ref{iull}},
A.~Claret\inst{\ref{iiaa}},
S.~Czesla\inst{\ref{ihamb}}, 
D.~Galadí\inst{\ref{icaha}},
Th.~Henning\inst{\ref{impia}}, 
S.~Khalafinejad\inst{\ref{ilza}}, 
M.~López-Puertas\inst{\ref{iiaa}},
D.~Montes\inst{\ref{iucm}},
A.~Quirrenbach\inst{\ref{ilza}}, 
I.~Ribas\inst{\ref{iice},\ref{iieec}},
M.~Stangret\inst{\ref{iiac},\ref{iull}},
F.~Yan\inst{\ref{iiag}}, 
M.\,R.~Zapatero~Osorio\inst{\ref{iastrb}},
M.~Zechmeister\inst{\ref{iiag}}
}

\institute{
\label{iiac} Instituto de Astrof\'isica de Canarias, 38200 La Laguna, Tenerife, Spain
\and
\label{iull} Departamento de Astrof\'isica, Universidad de La Laguna, 38206 La Laguna, Tenerife, Spain
\and
\label{iiag} Institut f\"ur Astrophysik, Georg-August Universit\"at G\"ottingen, Friedrich-Hund-Platz 1, 37077 G\"ottingen, Germany
\and
\label{iiaa} Instituto de Astrof\'isica de Andaluc\'ia (CSIC), Glorieta de la Astronom\'ia s/n, 18008 Granada, Spain
\and
\label{iice} Institut de Ci\`encies de l’Espai (ICE, CSIC), Campus UAB, c/ de Can Magrans s/n, 08193 Bellaterra, Barcelona, Spain
\and
\label{iieec}  Institut d’Estudis Espacials de Catalunya (IEEC), 08034 Barcelona, Spain
\and
\label{impia} Max Planck Institute for Astronomy, K\"onigstuhl 17, 69117 Heidelberg, Germany
\and
\label{ilza} Landessternwarte, Zentrum f\"ur Astronomie der Universit\"at Heidelberg, K\"onigstuhl 12, 69117 Heidelberg, Germany
\and
\label{icsh} University of Bern, Center for Space and Habitability, Gesellschaftsstrasse 6, CH-3012, Bern, Switzerland
\and
\label{ihamb} Hamburger Sternwarte, Gojenbergsweg 112, 21029 Hamburg, Germany
\and
\label{itaut} Th\"uringer Landessternwarte Tautenburg, Sternwarte 5, 07778 Tautenburg, Germany
\and
\label{iesac} Centro de Astrobiolog\'ia (CSIC-INTA), ESAC, Camino Bajo del Castillo s/n, 28692 Villanueva de la Ca\~nada, Madrid, Spain
\and
\label{icaha} Observatorio de Calar Alto, Sierra de los Filabres, 04550 Gérgal, Almería, Spain
\and
\label{iucm} Departamento de F\'isica de la Tierra y Astrof\'isica \& IPARCOS-UCM (Instituto de F\'isica de Part\'iculas y del Cosmos de la UCM), Facultad de Ciencias F\'isicas, Universidad Complutense de Madrid, 28040 Madrid, Spain
\and
\label{iastrb} Centro de Astrobiolog\'\i a (CSIC-INTA), Crta. Ajalvir km 4, Torrej\'on de Ardoz, Madrid, Spain
}

   \date{Received 19 August 2020 / Accepted dd Month 2020}
   
   \titlerunning{Chromatic RM of HD~189733b}
   \authorrunning{M. Oshagh, et al.}
   
   \abstract{Multiband photometric transit observations (spectro-photometric) have been used mostly so far to retrieve broadband transmission spectra of transiting exoplanets in order to study their atmospheres. An alternative method was proposed, and has only been used once, to recover broadband transmission spectra using chromatic Rossiter–McLaughlin observations. We use the chromatic Rossiter–McLaughlin technique on archival and new observational data obtained with the HARPS and CARMENES instruments to retrieve transmission spectra of HD~189733b. The combined results cover the widest retrieved broadband transmission spectrum of an exoplanet obtained from ground-based observation. Our retrieved spectrum in the visible wavelength range shows the signature of a hazy atmosphere, and also includes an indicateion for the presence of sodium and potassium. These findings all agree with previous studies. The combined visible and near-infrared transmission spectrum exhibits a strong steep slope that may have several origins, such as a super-Rayleigh slope in the atmosphere of HD~189733b, an unknown systematic instrumental offset between the visible and near-infrared, or a strong stellar activity contamination. The host star is indeed known to be very active and might easily generate spurious features in the retrieved transmission spectra. Using our CARMENES observations, we assessed this scenario and place an informative constraint on some properties of the active regions of HD~18973. We demonstrate that the presence of starspots on HD~189733 can easily explain our observed strong slope in the broadband transmission spectrum.}
   
   
   \maketitle

\section{Introduction}\label{sec:Introduction}

Transiting exoplanet atmospheres can be investigated in unprecedented detail through techniques such as transmission spectroscopy. This technique is in principle based on measuring the radius of a transiting exoplanet as a function of wavelength. Transmission spectroscopy has been established to be the most accessible technique for detecting broad- and narrow-band features generated by absorption or scattering of starlight by atoms, molecules, and particles in the planetary atmosphere \citep[e.g.,][]{Sing-10, Kreidberg-14, Wyttenbach-15, Sing-16, Mallonn-16, Lendl-17, Nikolov-18, Salz-18, Chen-18, Keles-19, Sanchez-Lopez-19}. Measurements of broadband features can only be acquired through multiband photometric transit (spectrophotometric) observations because the high-resolution spectroscopic technique
is insensitive to broad features. The most precise broadband transmission spectra have been obtained from space-based observations, mostly using the {\em Hubble Space Telescope} ({\em HST}) and the {\em Spitzer} telescope.

During its passage in front of its rotating host star, a transiting exoplanet creates a radial velocity (RV) signal, which is generated by obstructing the rotational velocity of the portion of stellar disk that is blocked by the planet. This is known as the Rossiter–McLaughlin (RM) effect \citep{Holt-1893, Rossiter-24, McLaughlin-24}. The RM signal contains several important pieces of information, including the sky-projected planetary spin-orbit angle \cite[a comprehensive review can be found in][and references therein]{Triaud-18}. Similar to the depth of the photometric transit light curve, the RM semiamplitude also scales with the radius of the transiting planet as

\begin{equation}
\mathit{A}_{RM} \simeq \frac{2}{3} \left(\frac{\mathit{R}_{p}^{2}}{\mathit{R}_{\star}^{2}} \right) \varv \sin \mathit{i}_{\star} \sqrt{1-\mathit{b}^{2}},
\end{equation}

\noindent where $\mathit{R}_{p}$ is the planetary radius, $\mathit{R}_{\star}$ is the host star radius, $\varv$ is the rotational velocity of the star, $\mathit{i}_{\star}$ is the host star inclination, and $\mathit{b}$ is the planet impact parameter \citep{Triaud-18}. 

Based on this fact, \citet{Snellen-04} developed a novel idea of retrieving the transmission spectra by measuring the RM signal amplitude in different wavelengths, the so-called chromatic RM.  \citet{Dreizler-09} later carried out a simulation study to examine the feasibility of this technique for different types of planets and host stars. While \citet{Snellen-04} advocated that chromatic RM can only be used to probe narrow-band features, such as sodium, \citet{DiGloria-15} employed this technique using the HARPS spectrograph data on HD~189733, and showed that it can be a powerful method for probing wide broadband features, which are challenging to be probed from the ground. 

Ground-based broadband transmission spectroscopy has been performed with the spectrophotometric method, which requires a reference star. On the other hand, the chromatic RM method does not need a reference star. The next generation of ground-based telescopes, such as the Extremely Large Telescope, will have a relatively small field of view that will pose limitations on accessing nearby reference stars. The chromatic RM method might therefore prove a promising technique for the characterization of planetary atmospheres.

\object{HD~189733b} is a hot Jupiter with a mass of 1.15\,$M_{\rm Jup}$ that
orbits a K dwarf every 2.2\,d. 
The system parameters are listed in Table~\ref{param}. HD~189733b is one of the most frequently observed and studied exoplanets, especially for its atmospheric signature through transmission spectra obtained from both ground- and space-based observations \citep{Sing-10, Pont-11, Sanchez-Lopez-19}.


We here proceed and apply the chromatic RM technique to retrieve a transmission spectrum of HD~189733b over a much wider wavelength range, using newly acquired CARMENES observations (both in the visible and near-infrared wavelength ranges). The paper is structured as follows. In Sect.~\ref{sec:Observation} we present the details of our observations and data reduction process. In Sect.~\ref{sec:Dataanalyis} we explainthe models that we used to analyze the observed chromatic RM signal. In Sect.~\ref{sec:ReproduceDiGloria} we examine whether the result obtained by \citet{DiGloria-15} can be reproduced using the exact same HARPS dataset. We present our retrieved broadband transmission spectrum of HD~189733b through HARPS and CARMENES observations in Sect.~\ref{sec:Result1} and interpret it in Sect.\ref{sec:atmospheremodel}. 
Next, Sect.\ref{sec:stellaractivity} is dedicated to assessing how the chromatic RM observations can be used to place constraints on the stellar active region properties, and also how much stellar activity contamination is expected in our retrieved transmission spectrum. We conclude our study
by summarizing the results in Sect.~\ref{sec:Conclusion}.

\section{Observations and data reduction}\label{sec:Observation}

\begin{table}
\caption{Stellar and planetary parameters of the HD~189733 system.}              
\centering                                      
\begin{tabular}{l c c c }          
\hline\hline                        
\noalign{\smallskip}
Parameter & Symbol & Unit & Value \\
\hline 
\noalign{\smallskip}
\noalign{\smallskip}
Orbital period & $P$ & d & 2.218573$^a$ \\
Scaled semimajor axis & $a/R_{\star}$ & ... & 8.715$^a$ \\
Orbital inclination & $i$ & deg & 85.508$^a$ \\
Spin-orbit angle & $\lambda$ & deg & --0.85$^a$ \\
Projected rotation velocity & $\varv \sin i$ & km\,s$^{-1}$ & 3.1$^a$ \\
\noalign{\smallskip}
\hline                                             
\end{tabular}
\begin{flushleft} 
$^a$ \citet{Triaud-09}\\
\label{param}
\end{flushleft}
\end{table}

\subsection{HARPS observations}

HARPS is a fiber-fed, cross-dispersed, high-resolution echelle spectrograph mounted at the ESO 3.6\,m telescope, and covers a wavelength range from 378 to 691\,nm \citep{Mayor-03}. \citet{DiGloria-15} used archival HARPS RV measurements during three transits of HD~189733b, taken on 7 September 2006, 19 July 2007, and 28 August 2007. We downloaded these observed spectra from the ESO archive for our analysis.

\subsection{CARMENES observations}

CARMENES is a high-resolution spectrograph installed at the 3.5\,m telescope at Calar Alto Observatory (Almer\'ia,
Spain). 
CARMENES consists of two cross-dispersed echelle spectrograph channels, which cover visible (CARMENES-VIS) wavelengths (520--960\,nm) and near-infrared (CARMENES-NIR) wavelengths (960--1710\,nm) \citep{Quirrenbach-14}. Two transits of HD~189733b were observed with CARMENES-VIS only on 8 August 2016 and 17 September 2016, and a third transit on 9 August 2019 was observed with CARMENES-VIS and CARMENES-NIR simultaneously. We summarize the details of each CARMENES observation in Table~\ref{tab:observationsummary}.

All raw spectra were processed using {\tt caracal} \citep{Caballero-16}. This code performs the basic spectral reduction process to obtain calibrated 1D spectra.

Wavelength calibration is crucial to obtain high-precision RVs, even when the spectrograph is stabilized. The CARMENES-VIS and CARMENES-NIR channels are equipped with Fabry-P\'erot etalons \citep{Seifert-12}, which were used for simultaneous drift measurements during the night \citep{Schafer-18}, and also to construct the daily wavelength solution \citep{Bauer-15}.

\begin{table*}
        \caption{Summary of CARMENES RM observations of HD~189733b.}
        \centering
        \begin{tabular}{lccc}
    \hline \hline 
    \noalign{\smallskip}
                Instrument & Night  & Number of spectra & Exposure time [s]\\
        \noalign{\smallskip}
        \hline
        \noalign{\smallskip}
                CARMENES-VIS & 8 August 2016 & 45 & 240\\
                CARMENES-VIS & 17 September 2016 & 57& 240\\
                CARMENES-VIS & 9 August 2019 & 85 & 240\\
                CARMENES-NIR & 9 August 2019 & 57 & 260\\
        \noalign{\smallskip}
        \hline
        \end{tabular}
        \label{tab:observationsummary}
\end{table*}

\subsubsection{Telluric correction}

CARMENES spectra, especially in the near-infrared, contain regions that are considerably contaminated by tellurics. To derive reliable high-precision RVs without masking a large number of pixels, we perform in Sect.~\ref{subsec:serval} and \ref{subsec:ccf} telluric line modeling based on {\molecfit} to correct the spectra as described by \citet{Bauer-20} and \citet{Nagel-20}.

\subsection{RV extraction from the spectra}

There are two main approaches to measuring RVs during the transit of an exoplanet and extracting the RM signal. 
One approach is based on the matching of each single spectrum with a reconstructed high signal-to-noise ratio (S/N) average template from all the observed spectra. This is known as the template-matching approach \citep{Butler-96, Guillem_TERRA}. The other apporach relies on a Gaussian fit to the cross-correlation function (CCF) of the observed spectra with a binary mask \citep{queloz1995spectroscopy, baranne1996elodie, Pepe-02, lafarga2020carmenesccf}. 
Each approach leads to a slightly different shape and amplitude of the RM signal, as was investigated theoretically by \citet{Boue-12}. In this study we examine and use both approaches.

\subsubsection{Template-matching: {\serval} }\label{subsec:serval}

In order to detect the slight wavelength-dependent amplitude changes in the chromatic RM, high-precision RV measurements have to be obtained. The challenge posed is to achieve m\,s$^{-1}$ precision in relatively narrow wavelength bins that only represent a fraction of the entire spectral range covered by the spectrograph. We used the code {\serval} \citep{Zechmeister2018A&A...609A..12Z}, which performs a least-square matching between individual spectra and the template, to compute the RVs. 

HD~189733 is known to be an active star, and its spot configuration is expected to change significantly between transit nights \citep{Boisse-09, Kohl-18}. Because changes in the line profiles are expected, combining spectra gathered at different epochs into a single template can therefore bias the derived RV measurements. However, the rotational period of HD~189733 is about 11.95\,d \citep{2008AJ....135...68H}, so that the spot configuration and the resulting line distortions are quasi-static during one transit. Furthermore, spectra taken during transit also show line distortions due to the RM effect, therefore these spectra were not used either to build the template. The cleanest way to measure RVs for the purpose of this work was to create a separate template for each observing night that consisted only of out-of-transit spectra. In doing so, we reduced the S/N of the template but avoided adding unrelated line distortions from the different nights (spot configurations). 

{\serval} measures RVs order by order, which typically results in 10\,nm wide bins. The RV precision achieved in one spectral order is insufficient to carry out atmospheric studies with the chromatic RM, however. We therefore combined the RVs derived from several single orders into 50\,nm bins (similar to the wavelength bins that were considered also in \citealp{DiGloria-15}). Because the number of lines available for RV extraction in the wavelength range covered by CARMENES-NIR was smaller \citep{Bauer-20}, we increased the range of the bins to 75--100\,nm. These RM curves are shown in Figs.~\ref{fig:HARPS-BESTFIT}, \ref{fig:CARMENES-VIS-BESTFIT}, and \ref{fig:CARMENES-NIR-BESTFIT}.



\subsubsection{Cross-correlation function: CCF}\label{subsec:ccf}

We also computed RVs following the CCF approach using the {\raccoon} code \citep{lafarga2020carmenesccf}. Instead of using a predefined mask, we created three weighted binary masks, one for each instrument, HARPS, CARMENES-VIS, and CARMENES-NIR, from the observations of HD~189733 themselves. To build each of the masks, we used the high S/N templates created by {\serval} to locate and select deep, narrow, and unblended absorption lines. The templates were built by coadding the out-of-transit observations available for each instrument.
The lines have an associated weight given by their contrast and inverse full width at half maximum (FWHM), as measured in the templates. 

We then used these masks to compute the CCFs of the observations. For the HARPS data, we also used two of the default masks used by the HARPS Data Reduction Software (DRS) created from spectral templates of a G2 and a K5 spectral type star. 

We computed a CCF for each individual order, and then combined them by coadding the CCFs corresponding to the same 50\,nm wide bins computed in Sect.~\ref{subsec:serval}. The CCF of each order was weighted according to the S/N and the number and quality of the lines used in the order. The final RV for each bin was obtained by fitting a Gaussian function to the coadded CCFs. Similar to the template-matching approach, we increased the wavelength bin size to 75--100\,nm in near-infrared.

\begin{table}
        \caption{Prior on free parameters }
        \centering
        \begin{tabular}{lc}
    \hline \hline 
        \noalign{\smallskip}
                Parameter & Prior \\
                \noalign{\smallskip}
                \hline
                \noalign{\smallskip}
        $R_{p}/R_{\star}$ & $\mathcal{N}(0.16;0.01)$\\
        Limb-darkening coefficients &  $\mathcal{N}(\texttt{LDTk};0.05)$\\
        $A_{GP} (m/s)$ &  $\mathcal{U}(0;10)$\\
        $\tau_{GP} (days)$ &  $\mathcal{N}(WL\tau_{GP};0.05)$\\
        \noalign{\smallskip}
                \hline
        \end{tabular}
        \label{tab:prior}
        \begin{flushleft} 
\textbf{Notes}: $\mathcal{U}(a;b)$ is a uniform prior with lower and upper limits of $a$ and $b$. $\mathcal{N}(\mu; \sigma)$ is a normal distribution with mean $\mu$ and width $\sigma$.
\end{flushleft}
\end{table}

To summarize, we obtained nine RM curves (either from {\serval} or CCF) for the
nine different passbands (each 50\,nm wide) for each of the three transits of HARPS and each of the three transits of CARMENES-VIS. We also obtained six RM curves, corresponding to the six different passbands (75--100\,nm wide) for a single transit in CARMENES-NIR.

\section{Analysis}\label{sec:Dataanalyis}

\subsection{Models for the Rossiter-McLaughlin effect}

To model the observed RM signal obtained through the CCF approach, we used the publicly available code \texttt{ARoME} \citep{Boue-12}. \texttt{ARoME} was developed and optimized to model the RM signals, which were extracted through the CCF-based approach. To model the RM observations obtained from the template-matching approach, with the {\serval} pipeline, we used the RM model based on the formulation from \citet{Ohta-05} and \citet{Hirano-11A}. This formulation was optimized for the RM signal obtained from the template-matching procedure. This model is implemented in the \texttt{PyAstronomy} Python package \citep{pya}.

Stellar activity can alter the out-of-transit stellar flux baseline in photometric transit light-curve observations, and the most efficient way to eliminate this effect is to normalize to the mean of the out-of-transit flux. In the RM observations, the active regions induce an offset and an additional underlying slope in the out-of-transit RV measurements (in addition to the gravitationally induced RV variation caused by the orbiting planet). The activity-induced out-of-transit RV slope can significantly differ from transit to transit because of variations in the configuration of stellar active regions over different nights, as shown by  \citet{DiGloria-15}, \citet{Oshagh-18}, and \citet{Boldt-20}. A conventional practical approach to eliminate this effect is to remove a linear trend from the out-of-transit RVs. A similar strategy was used by \citet{DiGloria-15}. We also removed the linear trend from each RM curve in each of the wavelength bins. 
In Sect.~\ref{sec:stellaractivity} we use the values of the removed slopes to estimate the properties of the stellar active regions. Subsequently, three transit observations (for HARPS and CARMENES-VIS) were combined for the modeling and the analysis.

\subsection{Gaussian process}

Several studies have shown that the photometric transit light curves of HD~189733b exhibited clear signatures of starspot occultation anomalies \citep[e.g.,][]{Sing-10,Pont-11}. The studies also found that these starspot occultations could have a significant impact on the accuracy of the planetary radius estimation \citep[e.g.,][]{Sing-10,Pont-11, Oshagh-14}. Because the physics and geometry behind the photometric transit light curve and RM effect are similar, they might be expected to be affected by the occultation of active regions in a similar way. To solve this problem, \citet{DiGloria-15} examined the residuals of the best-fit model to the white-light RM curve from HARPS, and searched for any sign of possible starspot-crossing events, or possible effects due to stellar differential rotation \citep{Albrecht-12b, Serrano-20}, convective blueshift and granulation \citep{Shporer-11, Cegla-16, Meunier-17}, or instrumental systematics. To eliminate this red noise from their individual RM curves, the residuals of white-light RM were subsequently removed from individual chromatic RMs in each wavelength bin. To properly account for this red noise, we decided to incorporate a Gaussian process (GP) model to our RM modeling, to perform a more robust fit and to obtain more accurate estimates.

Gaussian process is a general scheme for modeling correlated
noise \citep{Rasmussen-06}, and its power and advantages have been widely demonstrated in the field of exoplanetary research. For instance, it has been used to model and mitigate the jitter in RV time series  \citep[e.g.,][]{Haywood-14, Faria-16}, and also to correct the photometric transit observations \citep[e.g.,][]{Aigrain-16, Serrano-18}. Gaussian process has assisted in detecting small planetary signals embedded in stellar activity noise.  
 
To do this, we used the new implementation of GP in the \texttt{celerite} package \citep{Foreman-Mackey-17}, considering that some of the \texttt{celerite} kernels are well suited to describe different forms of the stellar activity noise. We selected the covariance as a Mat\'ern-3/2 kernel. To train our GP, we first fit an RM+GP model to the white-light RM signals (either from HARPS, CARMENES-VIS, or CARMENES-NIR). 

We modeled the observed chromatic RMs (obtained from either {\serval} or CCF) as the sum of the mean model (\texttt{PyAstronomy} or \texttt{ARoME}) and GP noise with Mat\'ern-3/2 covariance kernel. The posterior samples for our model were obtained through a Markov chain Monte Carlo (MCMC) using \texttt{emcee} \citep{Foreman-Mackey-13}. 

The prior on the GP timescale hyperparameter was controlled by a tight Gaussian around the best-fit value of the GP timescale obtained from the white-light RM fit. This is a reasonable assumption because it is expected that if the noise is generated by the stellar activity, it has a similar timescale in different wavelengths. However, the GP amplitude hyperparameter was controlled by an uninformative wide uniform prior. The prior on planet radius was controlled by Gaussian priors centered on the reported value from the white-light RM observation reported by \citet{Triaud-09} with a width according to the reported uncertainties ($\mathcal{N}(0.16;0.01)$). The priors on the limb-darkening coefficients\footnote{Depending on the model, in \texttt{ARoME} quadratic limb-darkening law is considered, and in \texttt{PyAstronomy} linear limb-darkening is taken into account.} were also constrained by Gaussian priors created using \texttt{LDTk} \citep{Parviainen-15} for all the wavelength bins. These priors are also listed in Table~\ref{tab:prior}. 

We derived the best-fit parameters and their associated uncertainties in our fitting procedure using an MCMC approach, using the affine invariant ensemble sampler \texttt{emcee} package. 
We randomly initiated the initial values for our free parameters for 30 MCMC chains inside the prior distributions. For each chain, we used a burn-in phase of 500 steps, and then again sampled the chains for 5000 steps.
Thus, the results concatenated to produce 150\,000 steps. We determined the best-fit values by calculating the median values of the posterior distributions for each parameter, based on the fact that the posterior distributions were Gaussian.

\section{Reproducing the results by \citet{DiGloria-15}}\label{sec:ReproduceDiGloria}

The most obvious verification is to examine if we can reproduce similar chromatic RM curves using the same HARPS observations as used by \citet{DiGloria-15} and if an identical transmission spectrum of HD~189733b can be constructed. 

\begin{figure}
        \centering
        \includegraphics[width=1.\linewidth]{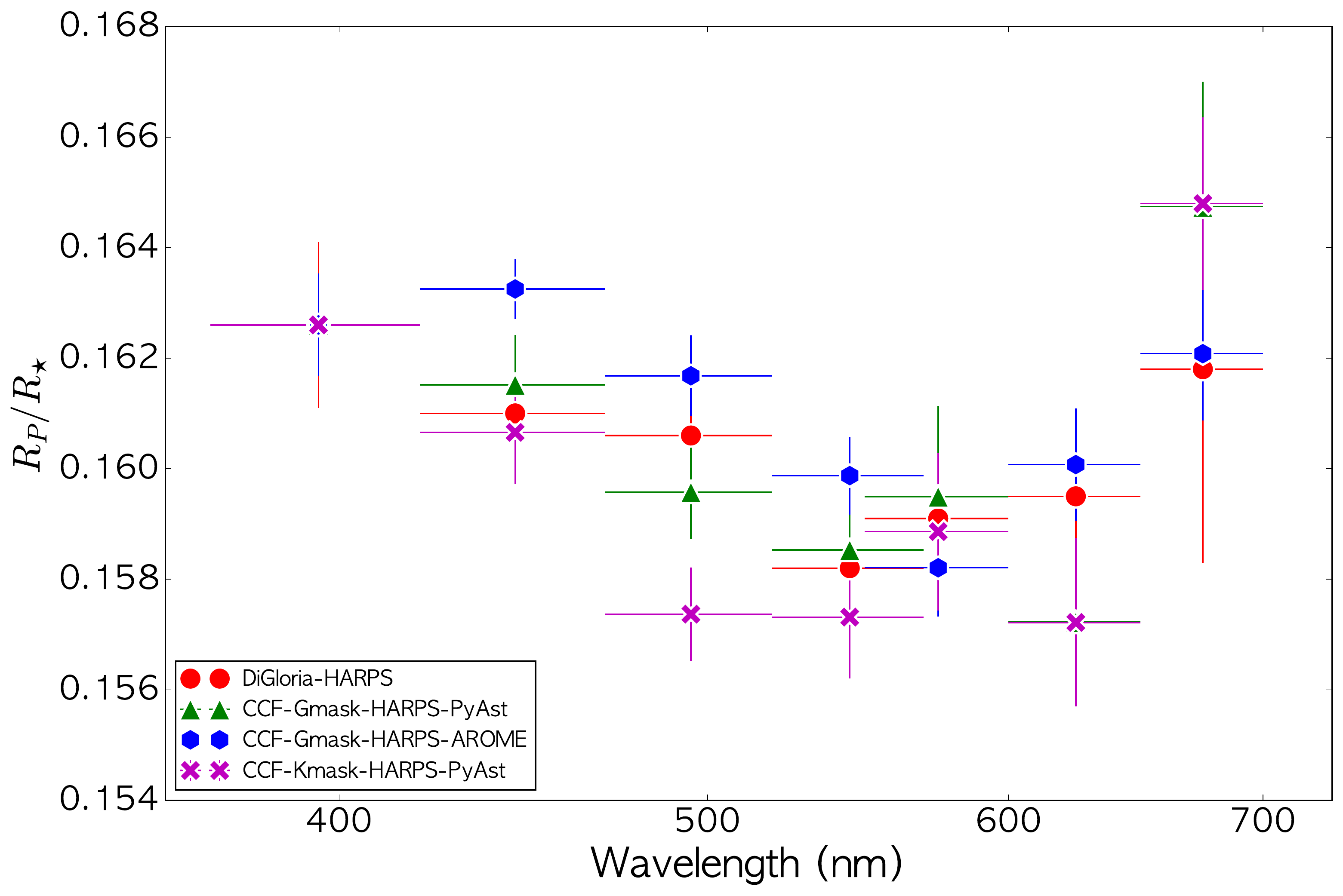}
        \caption{Comparison between the transmission spectra of HD~189733b obtained by \citet{DiGloria-15} (red circles), and the ones obtained from our analysis based on the CCF approach using two default binary mask of DRS, G2-mask (green triangles) and K5-mask (magenta crosses). We examined different RM modeling, such as \texttt{PyAstronomy} and \texttt{ARoME} on CCF approach using the G2 mask, displayed as green triangles and blue hexagons, respectively. All transmission spectra are offset to have the exact same value in the first bin, for the purpose of comparison.}
        \label{fig:Diffrentmask}
\end{figure}

\citet{DiGloria-15} used the CCF approach to derive RVs during the transits of HD~189733b, and used the G2 mask to generate CCFs in 50\,nm wavelength bins\footnote{Our host star is a K-type star, but \citet{DiGloria-15} decided to use the G2 mask.}. To repeat the exact same procedure, we also generated the CCFs in 50\,nm wavelength bins using the predefined G2 and K5 masks from the HARPS DRS. Then we fit the chromatic RMs in each wavelength bin using the GP+\texttt{PyAstronomy} and GP+\texttt{ARoME} models. The results are shown in Fig.~\ref{fig:Diffrentmask} and show a strong agreement between the retrieved transmission spectra of HD~189733b by \citet{DiGloria-15} and our CCF approach when the predefined masks are used. We also evaluated whether the use of different RM models implemented in the \texttt{PyAstronomy} and \texttt{ARoME} packages affect the retrieved transmission spectra. We found that the obtained transmission spectra are insensitive to and independent of the model, although the formulation by \citet{Ohta-05} used in \texttt{PyAstronomy} was not designed and optimized for the RM generated by the CCF technique approach.

\begin{figure}
        \centering
        \includegraphics[width=1.\linewidth]{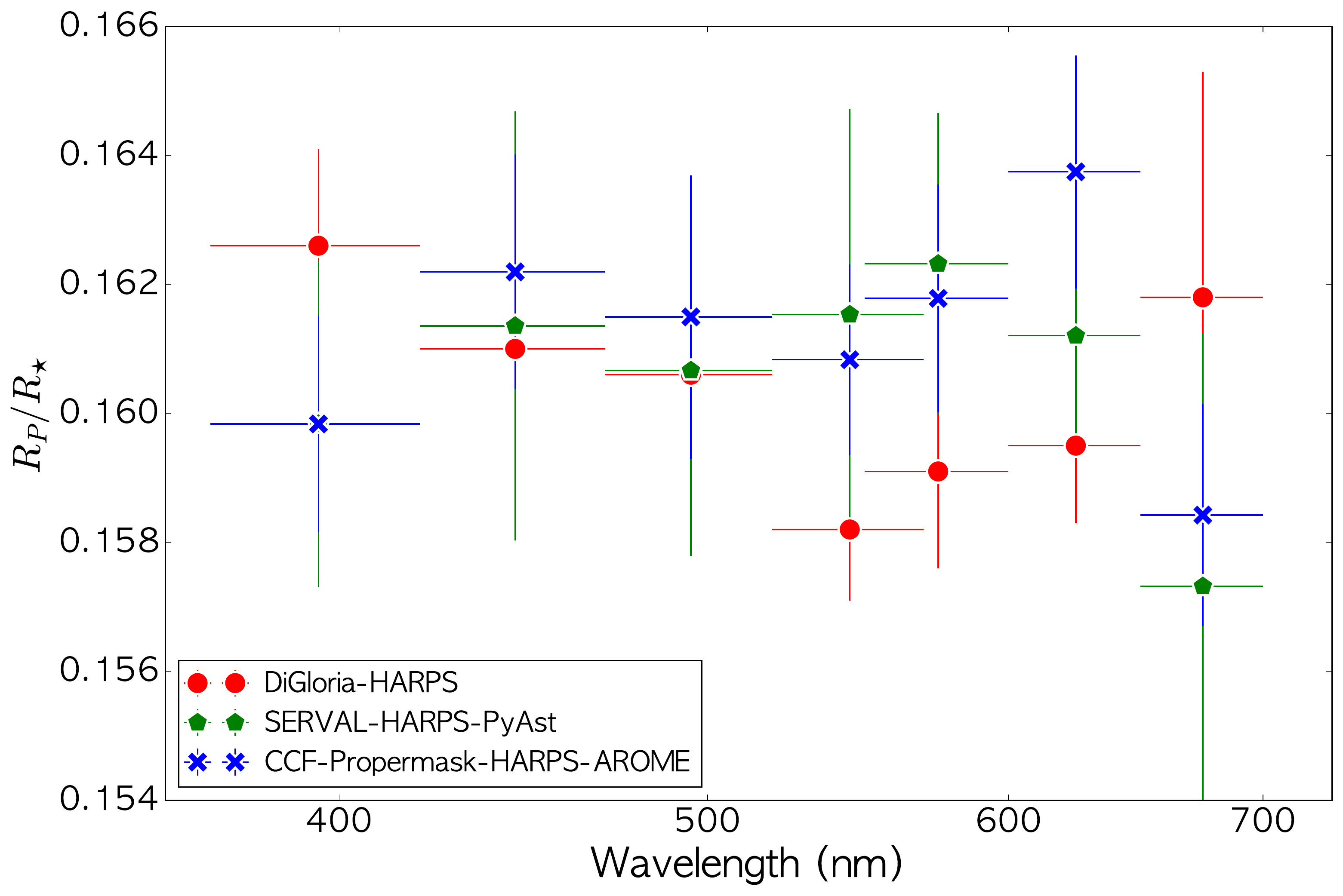}
        \caption{Comparison between the transmission spectra of HD~189733b obtained by \citet{DiGloria-15} (red circles) and those retrieved from our analysis based on the CCF approach using an adequate mask for HD~189733 (blue crosses) and from the template-matching approach using {\serval} (green pentagons).}
        \label{fig:SERVAL-Marina}
\end{figure}

Nevertheless, if the CCFs are generated using an adequate mask adopted specifically for HD~189733 (as was described in Sect.\ref{sec:Observation}), the retrieved transmission spectrum moderately deviates from the results by \citet{DiGloria-15}, as shown in Fig.~\ref{fig:SERVAL-Marina}. Moreover, when the template-matching approach is applied to extract RVs during transit (using the {\serval} pipeline, as explained in Sect.~\ref{sec:Observation}), the recovered transmission spectrum again marginally disagrees (statistically insignificant with a $p$-value of 0.3) with the reported one by \citet{DiGloria-15}. However, the {\serval} and adequate mask CCF strongly agree with each other. They also manifest a flatter transmission spectra, with much less increment in the planet radius toward shorter and longer wavelengths, in contrast to the findings from \citet{DiGloria-15}. 

There could be different explanations or causes for the conflicting results between an adequate mask and the DRS default binary masks. For instance, the blended spectral lines are considered in the default binary mask, whereas a careful selection of lines that neglected blended lines was performed for the adequate mask. A transiting planet induces a deformation (bump) in the spectral lines (RM signal is generated as a result of this bump), and the movement of the bump is different in the lines, which are blended. Other plausible explanation could be that each line has a different limb-darkening behavior \citep{Yan-15}, thus different mask with different lines can yield diverse shapes of RM curves.

\section{Results}\label{sec:Result1}

In this section we present the final retrieved broadband transmission spectrum through chromatic RM using newly acquired CARMENES observations combined with the HARPS observations. For both data sets we only used the {\serval} template-matching approach to generate RM curves during three transits in wavelength bins of 50\,nm in HARPS and CARMENE-VIS, and only for one transit in wavelength bins of 75--100\,nm in CARMENES-NIR. Then we used GP+\texttt{PyAstronomy} to fit the chromatic RM, and estimated the planet radius as a function of wavelength. The best-fit $R_{p}/R_{\star}$ ratio for the different wavelength bins is given in Table~\ref{tab:fitted Rp}, while the best-fit model to the RM in the individual passbands are shown in Figs.~\ref{fig:HARPS-BESTFIT}, \ref{fig:CARMENES-VIS-BESTFIT}, and \ref{fig:CARMENES-NIR-BESTFIT}.

\begin{figure}
        \centering
        \includegraphics[width=1.\linewidth]{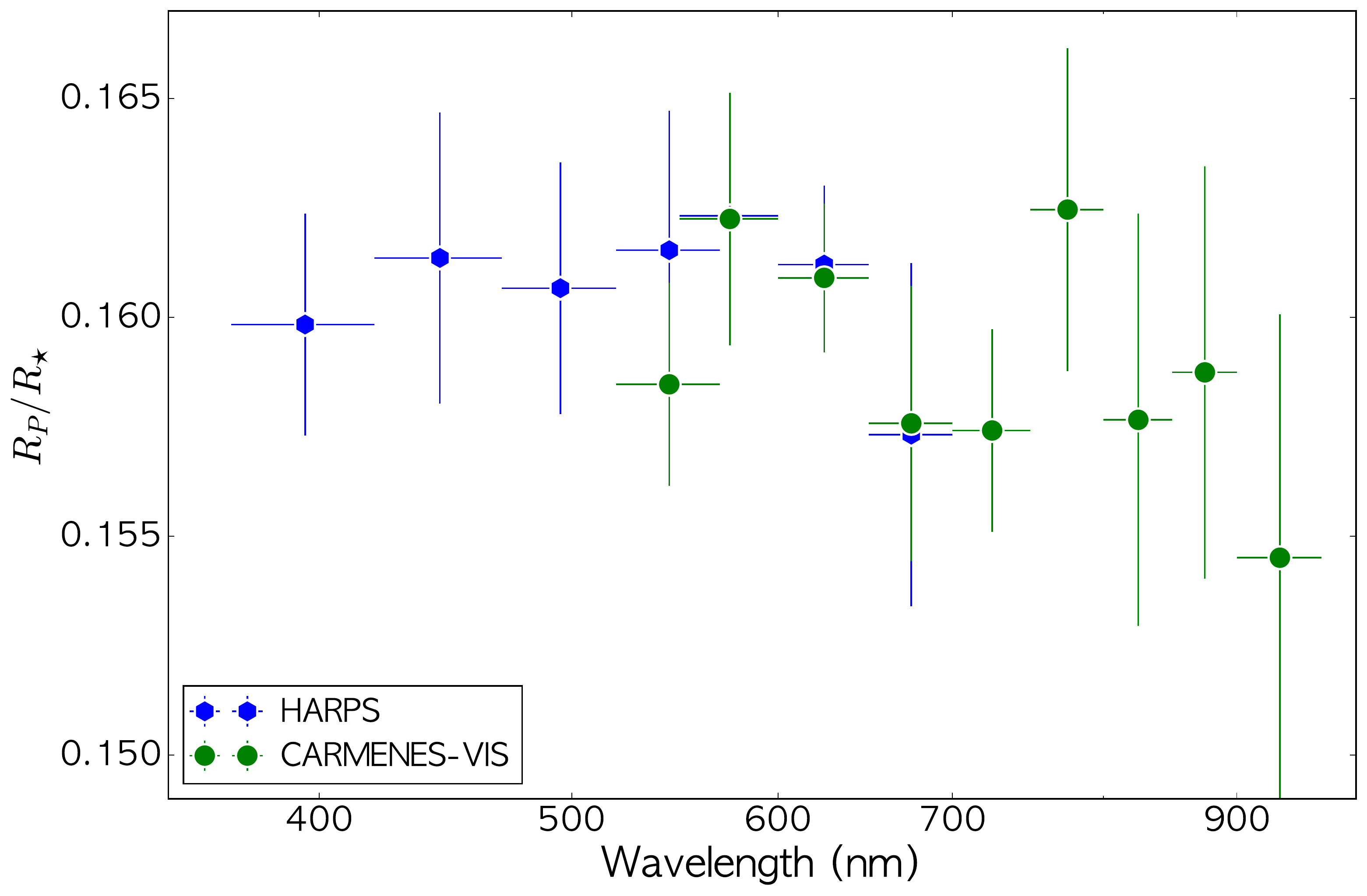}

        \centering
        \includegraphics[width=1.\linewidth]{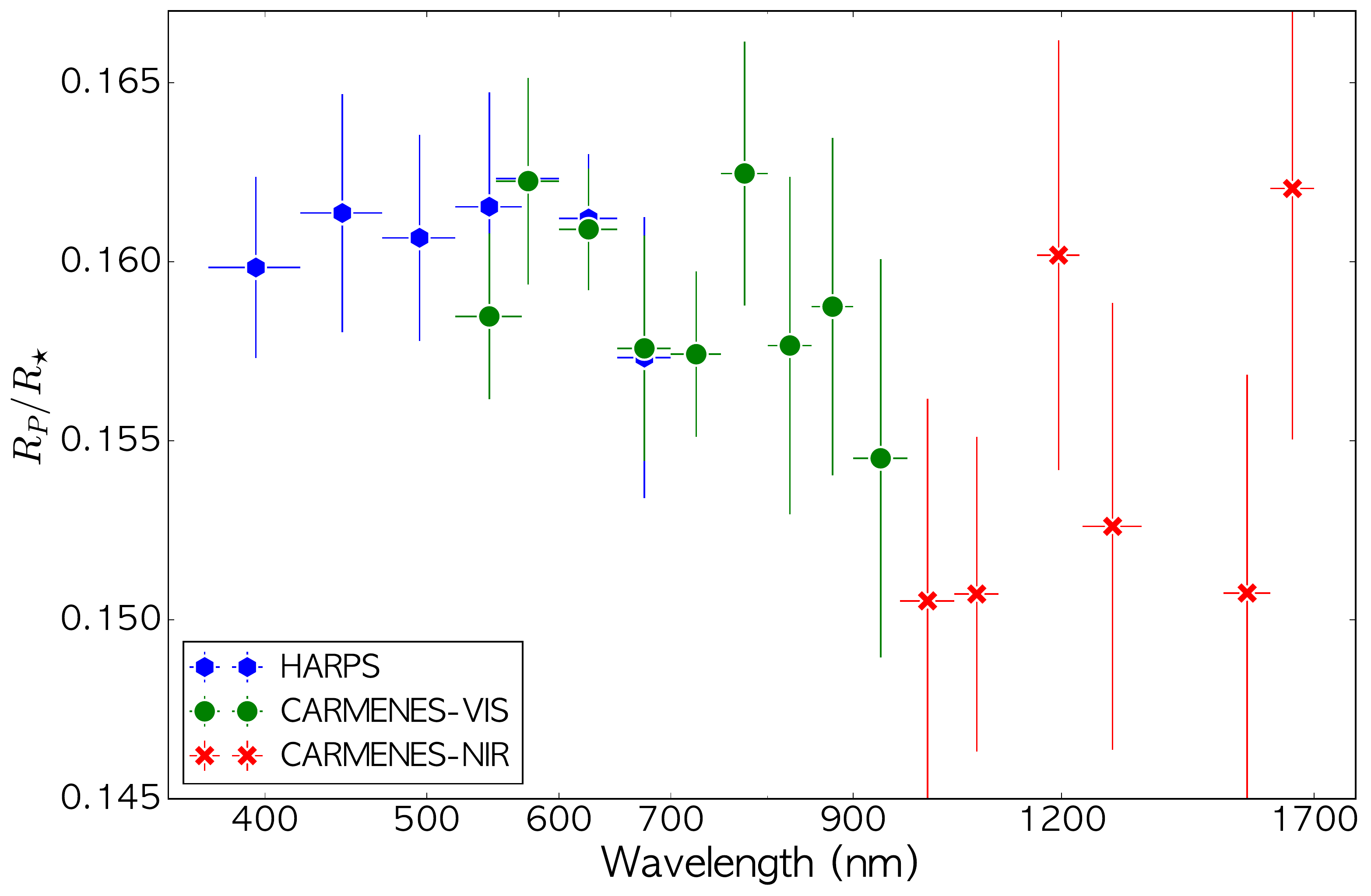}
        \caption{Top panel: Retrieved transmission spectra of HD~189733b obtained from HARPS (blue hexagons) and CARMENES-VIS (green circles). Bottom panel: Retrieved transmission spectra of HD~189733b obtained from HARPS, CARMENES-VIS, and CARMENES-NIR, shown with blue hexagons, green circles, and red crosses, respectively. All are obtained using chromatic RM through the template-matching approach and were fit with GP+\texttt{PyAstronomy}.}
        \label{fig:HARPS-CARM}
\end{figure}

One important and encouraging outcome is that the transmission spectra from CARMENES-VIS and HARPS agree well in overlapping wavelength regions, as shown in the top panel of Fig.~\ref{fig:HARPS-CARM}. The combined HARPS+CARMENES-VIS transmission spectrum exhibits a slope compatible with Rayleigh-scattering slope. It also shows some indication of the excess absorption from sodium and potassium. We explore this transmission spectrum in more detail in Sect.~\ref{sec:atmospheremodel}.


When the retrieved transmission spectra from CARMENES-NIR to HARPS+CARMENES-VIS are included, we recognize a steep drop between visible and near-infrared, and also some slight variation in the near-infrared, which is presented in the bottom panel of Fig.~\ref{fig:HARPS-CARM}. This distinct slope might also be interpreted as the Rayleigh-scattering slope. However, this slope is unrealistically steep. We probe this in more detail in Sect.~\ref{sec:atmospheremodel}. The second reason for this steeper slope could be related to the stellar activity. \citet{Boldt-20} demonstrated that stellar activity can easily mimic broadband features in transmission spectra retrieved from chromatic RM. We assess this possibility and also what can be learned from our observation about the stellar active region in Sect.~\ref{sec:stellaractivity}. The third reason might be the presence of an unknown systematic instrumental offset between the visible and near-infrared of CARMENES. However, evaluating this possibility requires a similar data set with simultaneous CARMENES-VIS and CARMENES-NIR RM observations, either for HD~189733b or any other transiting exoplanet, with simultaneous Fabry-P\'erot observations in both channels. Unfortunately, no such data set is available, and probing this scenario is therefore beyond scope of the current paper and will be pursued in a forthcoming publication.

\begin{table}
        \caption{Best-fit planetary radius derived in the different wavelength bins using observations from HARPS, CARMENES-VIS, CARMENES-NIR, and the combination of all instruments.}
        \centering
        \begin{tabular}{ccc}
    \hline \hline 
    \noalign{\smallskip}
                $\lambda$ [nm] & $R_{p}/R_{\star}$ & Error \\   
        \noalign{\smallskip}
    \hline
\noalign{\smallskip}
\multicolumn{3}{c}{\dotfill HARPS\dotfill}\\
\noalign{\smallskip}
395.0 & 0.1598  &  0.0025\\
445.0 &  0.1614  & 0.0032 \\
495.0 &  0.1607  &  0.0029\\
545.0 &  0.1615  &  0.0032\\
575.0 &  0.1623  &  0.0022 \\
625.0 & 0.1612  &  0.0018\\
675.0 &  0.1573  &  0.0036 \\
\noalign{\smallskip}
\multicolumn{3}{c}{\dotfill CARMENES-VIS\dotfill}\\
\noalign{\smallskip}
545.0 & 0.1585 &  0.0022 \\
575.0 &  0.1622 &  0.0029 \\
625.0 &  0.1609 &  0.0017 \\
675.0 &  0.1576 &   0.0029 \\
725.0 &  0.1574&   0.0021 \\
775.0 &   0.1625 &  0.0035 \\
825.0 &    0.1577 &     0.0047 \\
875.0 &   0.1587 &     0.0047\\
935.0 &    0.1545 &     0.0056\\
\noalign{\smallskip}
\multicolumn{3}{c}{\dotfill CARMENES-NIR\dotfill}\\
\noalign{\smallskip}
997.5 &    0.1499  &    0.0057\\
1067.5  &     0.1507 &     0.0044 \\
1195.0  &     0.1602   &    0.0060\\
1287.5  &     0.1526   &    0.0062\\
1550.0  &    0.1507    &   0.0061\\
1650.0  &     0.1620    &    0.0071\\
\noalign{\smallskip}
\multicolumn{3}{c}{\dotfill Combined\dotfill}\\
\noalign{\smallskip}
395.0  &    0.1598  &    0.0025\\
445.0 &    0.1614  &    0.0032\\
495.0 &    0.1607 &     0.0029\\
545.0 &     0.1599 &    0.0029\\
575.0 &     0.1622 &    0.0029 \\
625.0 &     0.1610  &    0.0011 \\
675.0 &     0.1574  &    0.0021\\
725.0 &     0.1574 &    0.0021 \\
775.0 &     0.1625  &     0.00347 \\
825.0 &     0.1577  &     0.0047\\
875.0 &     0.1587  &     0.0047\\
935.0 &     0.1545  &     0.0056\\ 
997.5 &     0.1499   &     0.0057\\
1067.5 &     0.1507   &   0.0044\\
1195.0 &     0.1602   &    0.0060\\
1287.5 &     0.1526   &    0.0062\\
1550.0 &     0.1507  &   0.0061\\
1650.0 &     0.1620   &   0.0071\\
        \noalign{\smallskip}
        \hline
        \end{tabular}
        \label{tab:fitted Rp}
\end{table}

\subsection{Atmospheric characterization}\label{sec:atmospheremodel}

\subsubsection{Forward-modeling}

We used \texttt{PLATON} \citep{Zhang-19} to perform forward-modeling of atmospheric properties of HD~189733b. \texttt{PLATON} is a fast, user-friendly, open-source code for retrieval and forward-modeling of exoplanet atmospheres written in Python. We fixed some of HD~189733b parameters, such as the isothermal temperature $T$, the planet radius, the C/O ratio, and atmospheric metallicity relative to the solar value $\log Z/Z_\odot$, to the known values from literature, and then varied the scattering slope. 

We introduced an offset into our HARPS+CARMENES-VIS of --0.0024 in $R_{p}/R_{\star}$ to match the transmission spectrum of HD~189733b obtained through {\em HST} multiband photometric transit observations \citep{Sing-16}. A similar shift was also introduced in \citet{DiGloria-15} to better match the {\em HST} observations. Because the absolute values of $R_{p}/R_{\star}$ estimated from RM curves can be affected by the choice of the stellar parameters, such as projected rotational velocity, and because we are moreover only interested in the relative changes of $R_{p}/R_{\star}$ as a function of wavelength, this arbitrary offset will not lead to any misinterpretation of the retrieved transmission spectra.

\begin{figure}
        \centering
        \includegraphics[width=1.\linewidth, height=6cm]{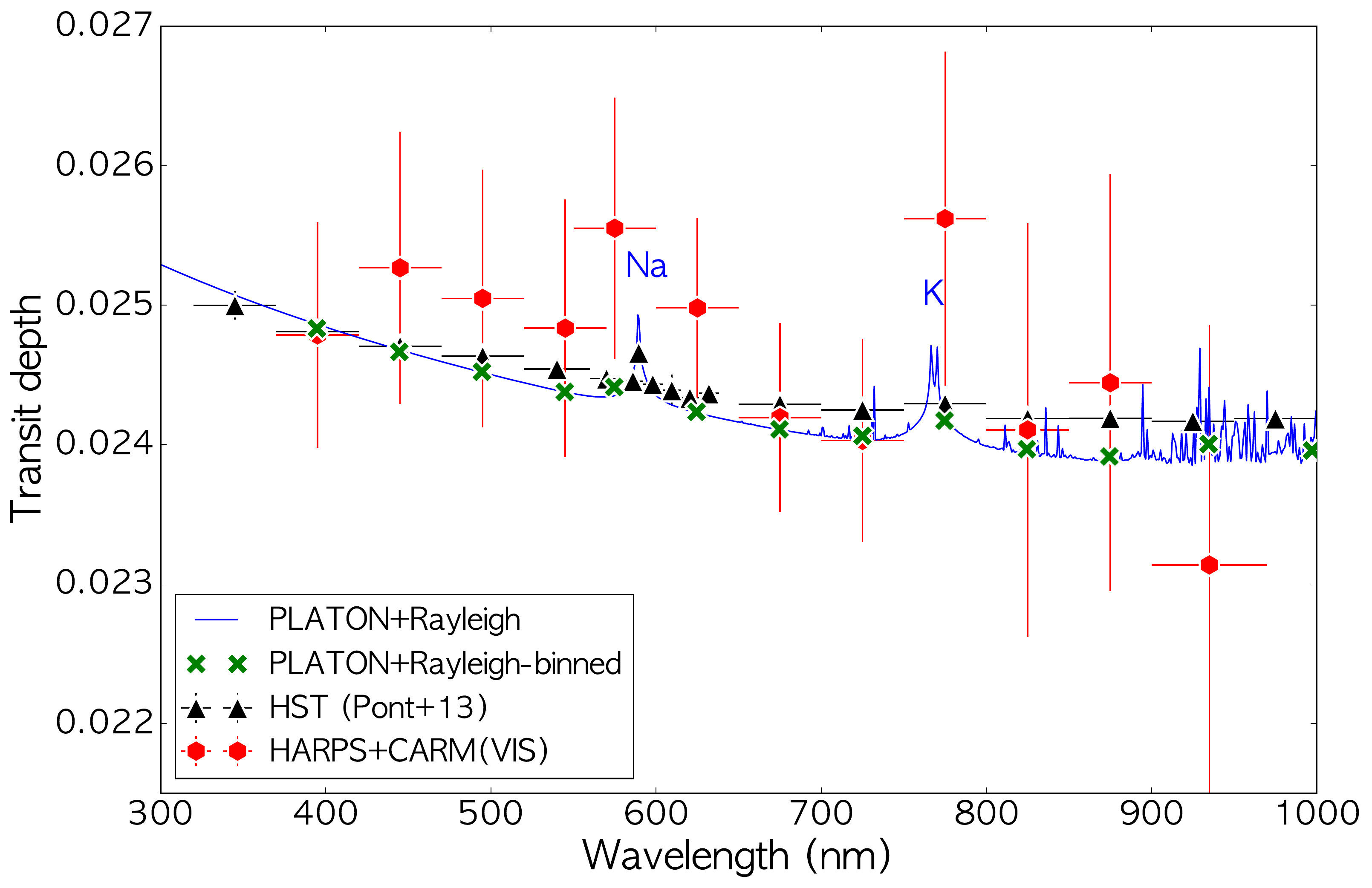}
        \caption{Retrieved broadband transmission spectra of HD~189733b obtained from combining HARPS and CARMENES-VIS (similar to the top panel of Fig.~\ref{fig:HARPS-CARM}) shown as red hexagons. The values for overlapping regions between HARPS and CARMENES-VIS were obtained by averaging the two values in each wavelength bin. The solid blue line represents the forward model obtained from \texttt{PLATON} considering a Rayleigh slope, and green crosses show the same \texttt{PLATON} model that is binned in the same wavelength bins as the observed ones. For comparison purposes, we also overplot the transmission spectra of HD~189733b obtained through multiband photometric transit observation obtained with {\em HST} as black triangles \citep{Pont-13}.}
        \label{fig:Forwardmodel-VIS}
\end{figure}

Our retrieved combined transmission spectrum from HARPS+CARMENES-VIS exhibits a slope that is consistent with a forward model with a Rayleigh slope of $\alpha=-9$, as was estimated in \citet{Pont-11}. It is shown in Fig.~\ref{fig:Forwardmodel-VIS}. The HARPS+CARMENES-VIS retrieved transmission spectrum also indicates a tentative indication of the presence of sodium (Na) and potassium (K) in the atmosphere of HD~189733b, which are again also present in the forward model, as shown in Fig.~\ref{fig:Forwardmodel-VIS}. Sodium has been detected in HD~189733b in several studies \citep[e.g.,][]{Wyttenbach-15}.
However, only a tentative detection was claimed for potassium, either from space-based {\em HST} observations \citep{Pont-13} or recently from ground-based observations using the PEPSI spectrograph \citep{Keles-19}. Our detection is only tentative, and should be taken with a grain of salt because our considered wavelength bins are quite wide and the uncertainties on the planetary radius estimates in each bin are large.


\begin{figure}
        \centering
        \includegraphics[width=1\linewidth, height=6cm]{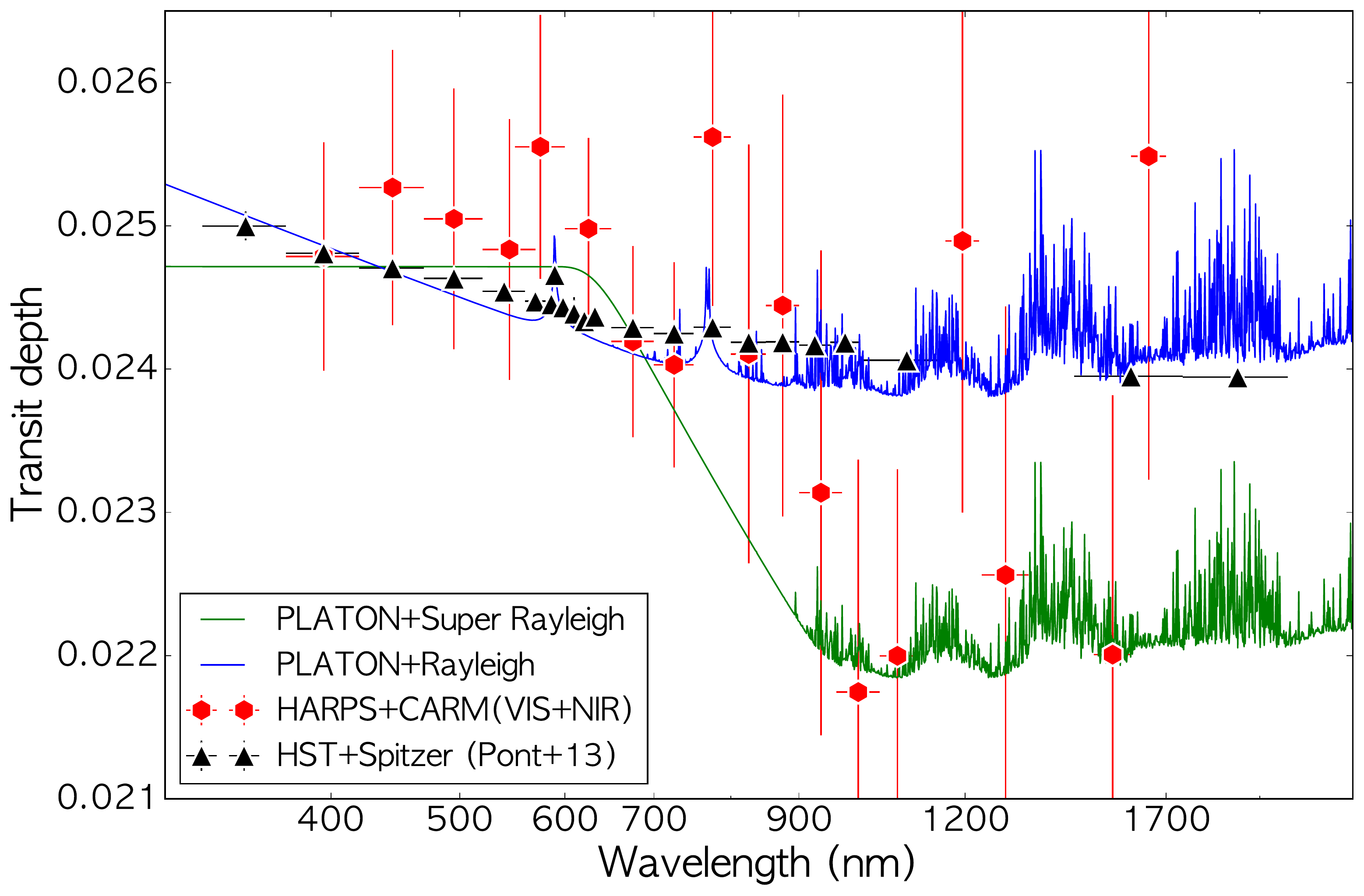}
        \caption{Retrieved broadband transmission spectra of HD~189733b obtained from combining HARPS, CARMENES-VIS, and CARMENES-NIR (similar to the bottom panel of Fig.~\ref{fig:HARPS-CARM}) shown as red hexagons. The solid blue line represents a forward model obtained from \texttt{PLATON} considering a Rayleigh slope. For comparison purposes, we also overplot the transmission spectra of HD~189733b obtained through multiband photometric transit observation obtained from {\em HST} and {\em Spitzer} as black triangles \citep{Pont-13}. The solid green line represents a forward model obtained from \texttt{PLATON} considering a super-Rayleigh slope. }
        \label{fig:Forwardmodel}
\end{figure}

We found that our HARPS+CARMENES-VIS+CARMENES-NIR transmission spectrum, as shown in Fig.~\ref{fig:Forwardmodel}\footnote{A similar negative offset of 0.0024
in $R_{p}/R_{\star}$ was also applied to the whole HARPS+CARMENES-VIS+CARMENES-NIR transmission spectrum, as we explained before.}, requires
a super-Rayleigh slope ($\alpha \ll -4$) to be described, which is at odds with the transmission spectra that were retrieved from {\em HST+Spitzer} observations and that suggested normal Rayleigh slope in the atmosphere of HD~189733b \citep{Pont-13}. A super-Rayleigh slope has been detected in other exoplanets, such as \object{HATS-8b} \citep{May-18}. \citet{Kazumasa-20} demonstrated that photochemical haze particles, which are formed in a vigorously mixing atmosphere, can produce a steep vertical opacity gradient that can lead to a super-Rayleigh slope. However, the atmospheric temperature of HD~189733b (equilibrium temperature $ \sim$ 1100\,K) is too high to sustain hydrocarbon hazes. In Sect.~\ref{sec:stellaractivity} we probe in detail how much the stellar activity might contribute to the observed slope.

\subsubsection{Inverse modeling}

We used {\tt petitRADTRANS} ({\tt pRT}) \citep{molliere_petitradtrans:_2019} to perform the retrieval. We assumed a Guillot profile for the temperature structure \citep{guillot_radiative_2010}, a parameterized Rayleigh slope, gray cloud deck, and three species (H$_2$O, Na, and K) in the model atmosphere. We excluded CARMENES-NIR and only used HARPS+CARMENES-VIS data for the reasons discussed in the previous section and because of the large uncertainties in the CARMENES-NIR.

\begin{figure}
        \centering
        \includegraphics[width=1.\linewidth]{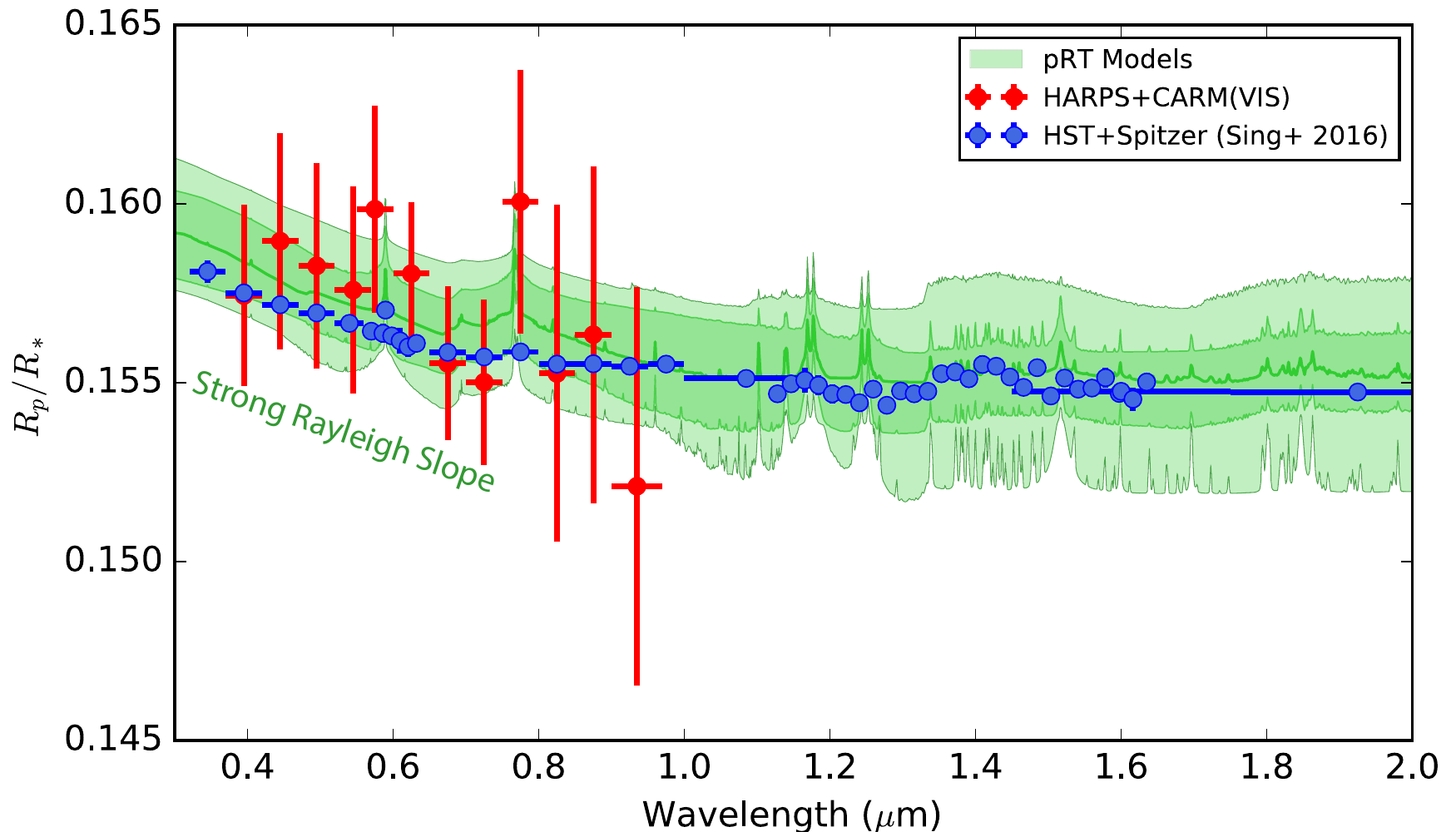}
        \caption{Atmospheric fitting of HD~189733b with {\tt pRT}. For the model spectra, the dark green area corresponds to the region of the posteriors between the 16\,\% and 84\,\% quantiles, and the light green area shows the region between the 1\,\% and 99\,\% quantiles. }
        \label{fig:retrieval}
\end{figure}

The best-fit model and the confidence intervals are shown in Fig.~\ref{fig:retrieval}\footnote{A similar negative offset of 0.0024
in $R_{p}/R_{\star}$ was applied here also, as we explained before.}. The estimated posteriors are presented in the appendix (Fig.~\ref{fig:retrieval_corner}). We found an agreement between the models and the results from \citet{Sing-16}, which indicates consistency between our measurements and those of the {\em HST}, even at wavelengths beyond CARMENES-VIS measurements. Our models suggest tentative evidence of Na and K presence, although their abundances are not constrained because of the large uncertainties in the HARPS+CARMENES-VIS combined spectrum. In particular, the abundance of potassium appears to be overestimated given the measurements at around 800\,nm, which might be caused by stellar activity \citep{Rackham-17}.

HD~189733b is expected to be a cloudy exoplanet given its temperature and composition \citep[e.g.,][, and references therein]{molaverdikhani2020role}, and our retrieved cloud-deck pressure level of around 100\,mbar agrees with this scenario. However, observations of molecular features at longer wavelengths are required to constrain the cloudiness of this planet. A distinct Rayleigh slope of $\alpha \sim -10$ was also retrieved and is consistent with the \citet{Sing-16} finding in the optical, as discussed in the previous section.

\subsection{Stellar activity contamination}\label{sec:stellaractivity}

The transmission spectra retrieval relies vigorously on accurate knowledge of the host star spectrum. This means that stellar activity can contaminate the retrieved transmission spectra. Several studies have explored the effect of stellar active regions and found that it could imitate broadband features and thus affect the strength of narrow-band features \citep[both atomic and molecular --][]{Oshagh-14,McCullough-14, Scandariato-15, Barstow-15, Herrero-16, Rackham-17, Rackham-19, Cauley-18, Mallonn-18, Tinetti-18, Apai-18}. There have been some inconsistent results observationally. For instance, \citet{Sedaghati-17} reported the first detection of TiO in the atmosphere of WASP-19b, which transits a very active star.
However, \citet{Espinoza-19} did not detect any sign of TiO absorption through new independent multiband photometric observations.


In the RM observations, the active regions induce an offset and an additional underlying slope in the out-of-transit RV measurements (in addition to the gravitationally induced RV variation induced by the orbiting planet). The activity slope can significantly differ from transit to transit as a result of the variation in the configuration of stellar active regions in different nights, as shown by \citet{Oshagh-18} and \citet{Boldt-20}. A conventional practical approach to eliminate this offset and slope is to remove a linear trend from the out-of-transit RVs. 

To examine how much our retrieved transmission spectrum from chromatic RM is affected by the stellar activity, we studied the RV slope of out-of-transit in chromatic RMs in detail. For this we considered the only transit of HD~189733 that was obtained with both CARMENES-VIS and CARMENES-NIR on 9 August 2019. There are two reasons for selecting this data set. First, because this is the only data set that covers a wide wavelength range simultaneously. The second reason is that combining observations from several nights can mean combining contributions of several unrelated active region configuration, which might lead to mixing signals and might blur the whole picture. 

\begin{figure}
        \centering
        \includegraphics[width=1.\linewidth]{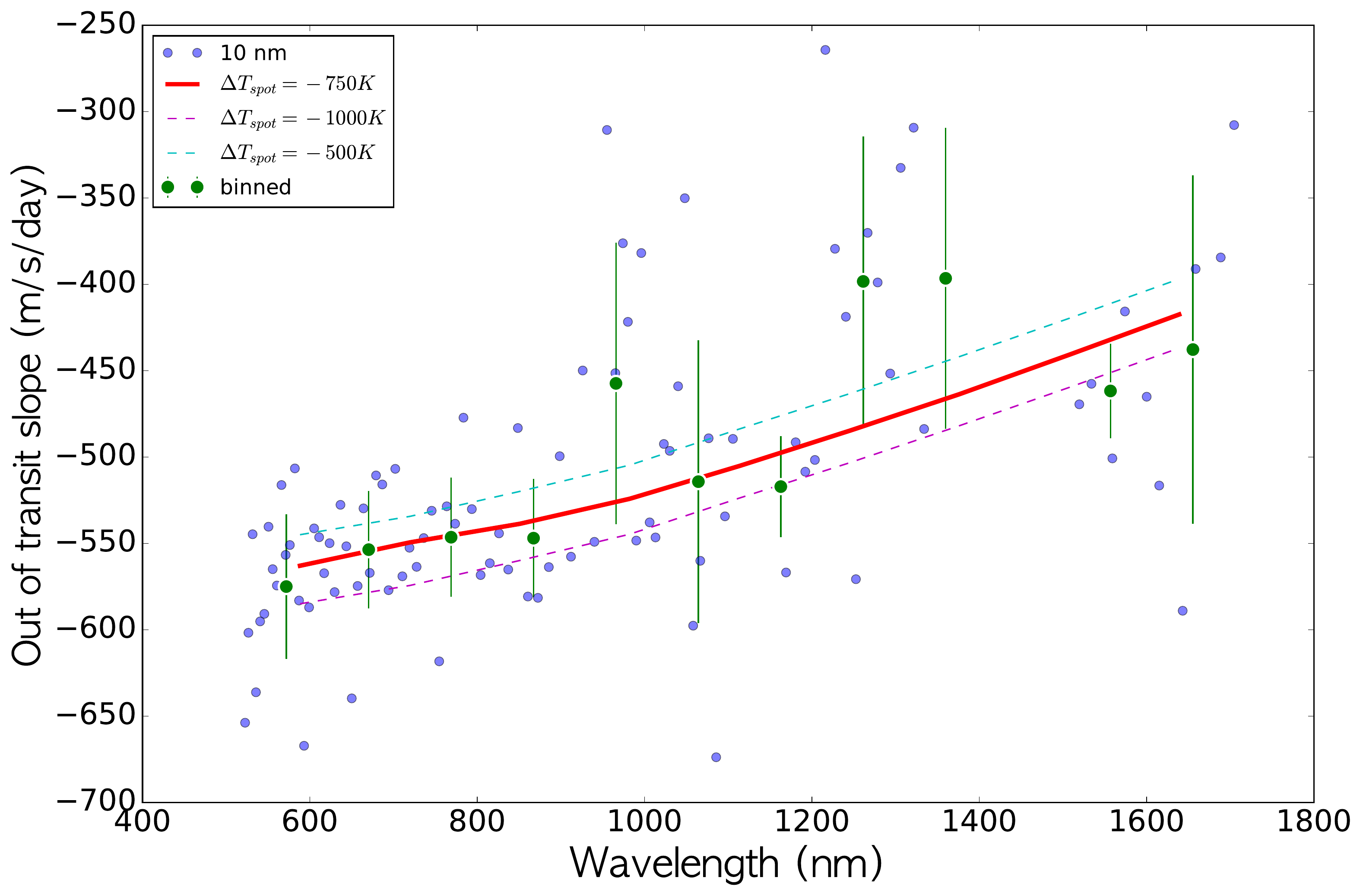}
        \caption{Estimated slopes of out-of-transit chromatic RM of one transit of HD~189733b observed with CARMENES (VIS+NIR) in 10\,nm and 50\,nm wavelength bins, represented as blue and green dots, respectively. The best-fit model of \texttt{SOAP3.0} with a single spot with a temperature difference of 750\,K cooler than the HD~189733 photosphere is indicated by the solid red line. The dashed green and purple lines show similar \texttt{SOAP3.0} models with +250\,K and --250\,K temperature differences, respectively.}
        \label{fig:FITSLOPE}
\end{figure}

For this analysis we also recalculated the RV time series during that night using {\serval}, but this time in 10\,nm wavelength bins (instead of 50\,nm in visible or 75--100\,nm in near-infrared as in Sect.~\ref{sec:Observation}) and fit a linear trend to the out-of-transit of each RM (both in 10\,nm and also 50--100\,nm passbands). We present the best-fit value of the slope in each wavelength bin in Fig.~\ref{fig:FITSLOPE}. We also show the best-fit linear model to 50--00\,nm passband RMs in Fig.~\ref{fig:SLOPE}. As Fig.~\ref{fig:FITSLOPE} clearly shows, the slope of the out-of-transit becomes less steep at longer wavelengths. This is unambiguous evidence that these slopes have been generated by active regions, whose temperature contrast is also wavelength dependent and becomes weaker at longer wavelengths. 

To place constraints on the characteristics of active regions that generated this trend, we used the publicly available tool \texttt{SOAP3.0}, which uses a pixellation method to simulate a transiting planet in front of a rotating star that harbors a different number and type of active regions, and delivers the photometric and RV measurements of the system \citep{Boisse-12, Dumusque-14, Oshagh-13a, Akinsanmi-18}. \texttt{SOAP3.0} does not only take the flux contrast effect of the stellar active regions into consideration, but also includes the RV shift caused by the inhibition of the convective blueshift within those regions.
Because the latest version of \texttt{SOAP3.0} performs the simulations at one single wavelength, we modified the code to be able to adjust the wavelength as one of the input parameters \citep{Boldt-20}. After adjusting \texttt{SOAP3.0}, the code automatically takes care of wavelength-dependent parameters, such as the active region contrast (based on Planck's law) and coefficients of the quadratic stellar limb-darkening law ($u_{1}$ and $u_{2}$), which are adopted from the \texttt{LDTk} model \citep{Parviainen-15}.

We assumed that HD~189733 harbors a spot with a filling factor of 4\,\%, which is a reasonable size because similarly sized spots were detected using spot-crossing anomalies analysis \citep{Sing-10, Herrero-16}. During our fitting procedure, we allowed the temperature difference between the spot and photosphere to vary as our only free parameter, whereas the other parameters were adjusted to the stellar and planetary parameters reported in Table~\ref{param}. We obtained the best-fit temperature difference between the spot and photosphere by minimizing the reduced $\chi^{2}$. The best-fit estimated that the spot temperature should be around 750\,K cooler than its surrounding photosphere, as shown in Fig.~\ref{fig:FITSLOPE}. This estimate concurs with the spot temperature contrasts that were previously estimated for this star \citep{Sing-10, Pont-11, Mancini-17}.

As was shown in a simulation study by \citet{Boldt-20}, such an active region (with a filling factor of 4\,\% and a temperature difference of 750\,K between the spot and photosphere) can easily mimic strong spurious broadband features, such as a strong Rayleigh scattering slope with up to 20\,\% variation of $R_{p}/R_{\star}$ in the retrieved transmission spectra from chromatic RM. However, the same study suggested that the probability of mimicking strong broadband features becomes lower when several RM observations are acquired during several transits and are combined, assuming that the stellar active regions evolve and disappear from transit to transit. Our retrieved transmission spectra in Sect.~\ref{sec:Result1} were achieved by combining six transit observations (three transits with HARPS and three transits with CARMENES). It might therefore be speculated that the contamination from stellar activity of the retrieved transmission spectrum should be eliminated and apparent features should have emerged from the atmosphere of HD~189733b. However, HD~189733 exhibits large active regions, and large active regions commonly live longer. Several studies have found evidence for the existence of long-lived spots on HD~189733 \citep{Boisse-09, Herrero-16} that can persist during all our transit observations, and therefore our retrieved transmission spectrum can still be contaminated by stellar activity.

Finally, to precisely evaluate the amplitude of stellar activity contamination imprinted on our retrieved transmission spectra, we performed a simulation test. We simulated with \texttt{SOAP3.0} a mock chromatic RM observation of an atmosphere-less planet with parameters similar to HD~189733b. We would like to emphasize that the transmission spectrum of an atmosphere-less exoplanet should in principle be flat without any broad- or narrow-band features. In our simulation we considered a starspot on the surface of the host star with properties (the spot filling factor and temperature contrast) equal to what was estimated above. Because we were interested in quantifying the maximum effect of starspots on the chromatic RM, the spot longitude was adjusted to be at the center of the stellar disk during the transits. We fit our mock chromatic RMs to estimate the planetary radius in each wavelength, similar to the simulations performed by \citet{Boldt-20}.

\begin{figure}
        \centering
        \includegraphics[width=1.\linewidth]{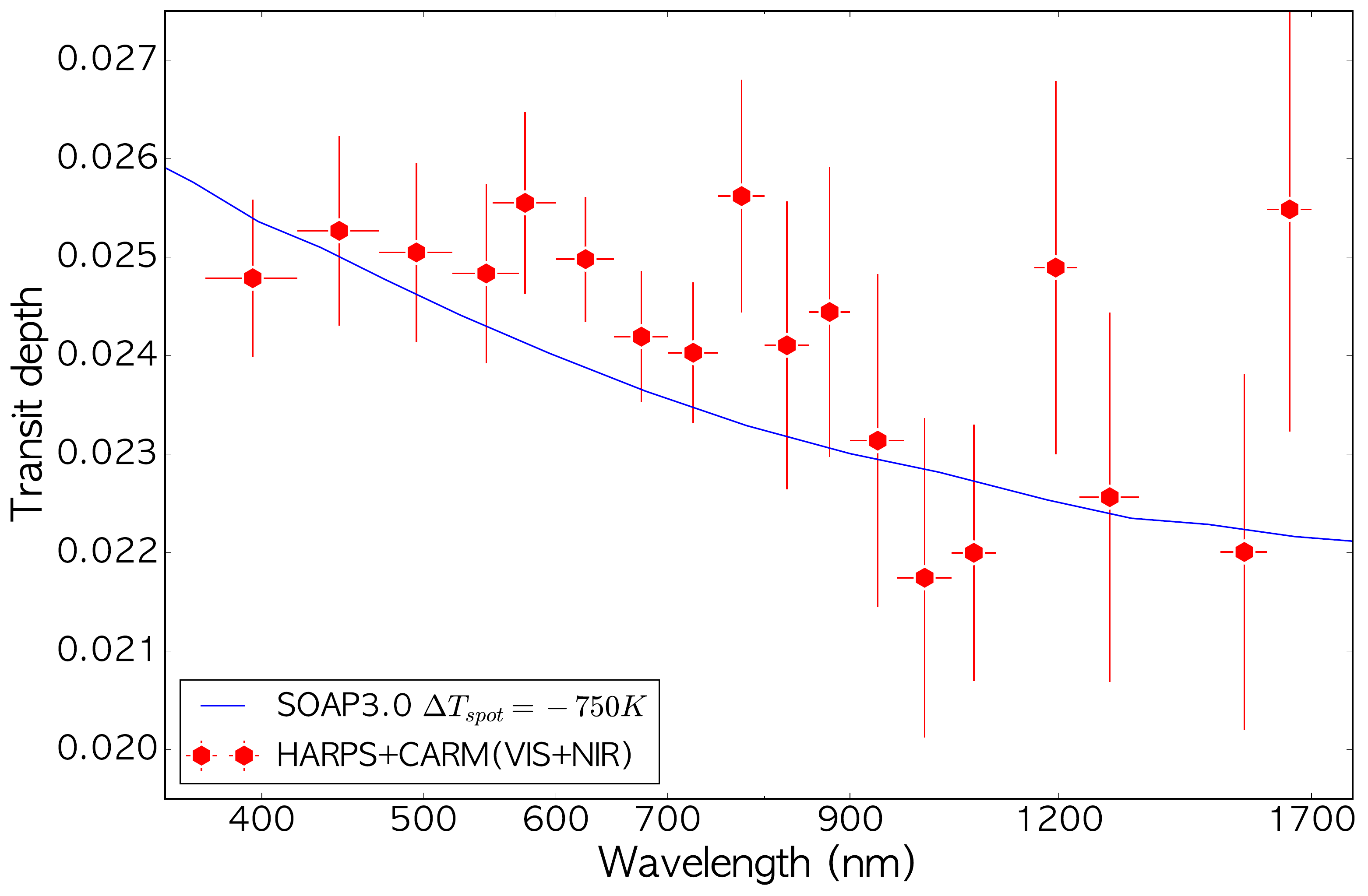}
        \caption{Retrieved transmission spectrum of an atmosphere-less planet obtained through \texttt{SOAP3.0} simulated chromatic RMs, considering a starspot with the properties estimated in Fig.~\ref{fig:FITSLOPE} (solid blue line). Our observed transmission spectrum from HARPS+CARMENES (VIS+NIR) is also shown as red hexagons.}
        \label{fig:SOAP}
\end{figure}

Our retrieved transmission spectrum through the simulated chromatic RMs is shown in Fig.~\ref{fig:SOAP}. This result indicates that the steep slope observed in our broadband transmission spectrum between the visible and near-infrared could indeed be generated by stellar activity.

\section{Conclusions}\label{sec:Conclusion}

We used a novel technique of chromatic Rossiter–McLaughlin observations on archival and new observations obtained from HARPS and CARMENES instruments to retrieve the transmission spectrum of HD~189733b. We found that if the CCF approach is applied to extract the RVs during the transit, then it is necessary to use a tailored mask that is specifically adapted to the host star spectrum. Moreover, the retrieved transmission spectrum from the adequate mask agrees with the spectrum from the template-matching approach. 

Our retrieved transmission spectrum through chromatic RM of the HARPS observations yields a similar spectrum to that from CARMENES-VIS observations, especially in the wavelength region where HARPS and CARMENES-VIS overlap. The combined HARPS and CARMENES-VIS transmission spectrum in the visible range exhibits a slope compatible with a Rayleigh-scattering slope, and it also indicates excess absorption from sodium and potassium. This agrees with previous studies.

The combined transmission spectra from HARPS, CARMENES-VIS, and CARMENES-NIR cover the widest retrieved broadband transmission spectrum of an exoplanet obtained from ground-based observations. However, visible and near-infrared transmission spectra exhibit an exceptionally strong slope that might have several origins, such as a super-Rayleigh slope in the atmosphere of HD~189733b, an unknown systematic instrumental offset between the visible and near-infrared, or most likely, stellar activity contamination. 

The host star is indeed known to be active, and this might easily mimic spurious features in the retrieved transmission spectra. Using our CARMENES observation, we placed constraints on the temperature contrast of the starspot on HD~18973. We also demonstrated that this starspot can easily generate the observed strong slope in the broadband transmission spectrum.

\begin{acknowledgements}
M.O. acknowledges the support of the Deutsche
Forschungs\-gemeinschaft (DFG) priority program SPP 1992 ``Exploring the Diversity of Extrasolar Planets (RE 1664/17-1)''. 
CARMENES is an instrument for the Centro Astron\'omico Hispano-Alem\'an (CAHA) at Calar Alto (Almer\'{\i}a, Spain), operated jointly by the Junta de Andaluc\'ia and the Instituto de Astrof\'isica de Andaluc\'ia (CSIC). CARMENES was funded by the Max-Planck-Gesellschaft (MPG), the Consejo Superior de Investigaciones Cient\'{\i}ficas (CSIC), the Ministerio de Econom\'ia y Competitividad (MINECO) and the European Regional Development Fund (ERDF) through projects FICTS-2011-02, ICTS-2017-07-CAHA-4, and CAHA16-CE-3978, 
and the members of the CARMENES Consortium 
  (Max-Planck-Institut f\"ur Astronomie,
  Instituto de Astrof\'{\i}sica de Andaluc\'{\i}a,
  Landessternwarte K\"onigstuhl,
  Institut de Ci\`encies de l'Espai,
  Institut f\"ur Astrophysik G\"ottingen,
  Universidad Complutense de Madrid,
  Th\"uringer Landessternwarte Tautenburg,
  Instituto de Astrof\'{\i}sica de Canarias,
  Hamburger Sternwarte,
  Centro de Astrobiolog\'{\i}a and
  Centro Astron\'omico Hispano-Alem\'an), 
  with additional contributions by the MINECO, 
  the DFG through the Major Research Instrumentation Programme and Research Unit FOR2544 ``Blue Planets around Red Stars'', 
  the Klaus Tschira Stiftung, 
  the states of Baden-W\"urttemberg and Niedersachsen, 
  and by the Junta de Andaluc\'{\i}a. 
  Based on data from the CARMENES data archive at CAB (CSIC-INTA). 
  We acknowledge financial support from the Agencia Estatal de Investigaci\'on of the Ministerio de Ciencia, Innovaci\'on y Universidades and the ERDF  through projects   PID2019-109522GB-C51/2/3/4   
  PGC2018-098153-B-C33          
  AYA2016-79425-C3-1/2/3-P,     
  ESP2016-80435-C2-1-R,         
and the Centre of Excellence ``Severo Ochoa'' and ``Mar\'ia de Maeztu'' awards to the Instituto de Astrof\'isica de Canarias (SEV-2015-0548), Instituto de Astrof\'isica de Andaluc\'ia (SEV-2017-0709), and Centro de Astrobiolog\'ia (MDM-2017-0737), and the Generalitat de Catalunya/CERCA programme. We thank the anonymous referee for
insightful suggestions, which added the clarity of this paper.

\end{acknowledgements}

\bibliographystyle{aa} 
\bibliography{mah_fixed}

\appendix
\section{Best-fit RM curves}

\begin{figure*}
 \subfloat{\includegraphics[width=0.32\textwidth, height=4 cm]{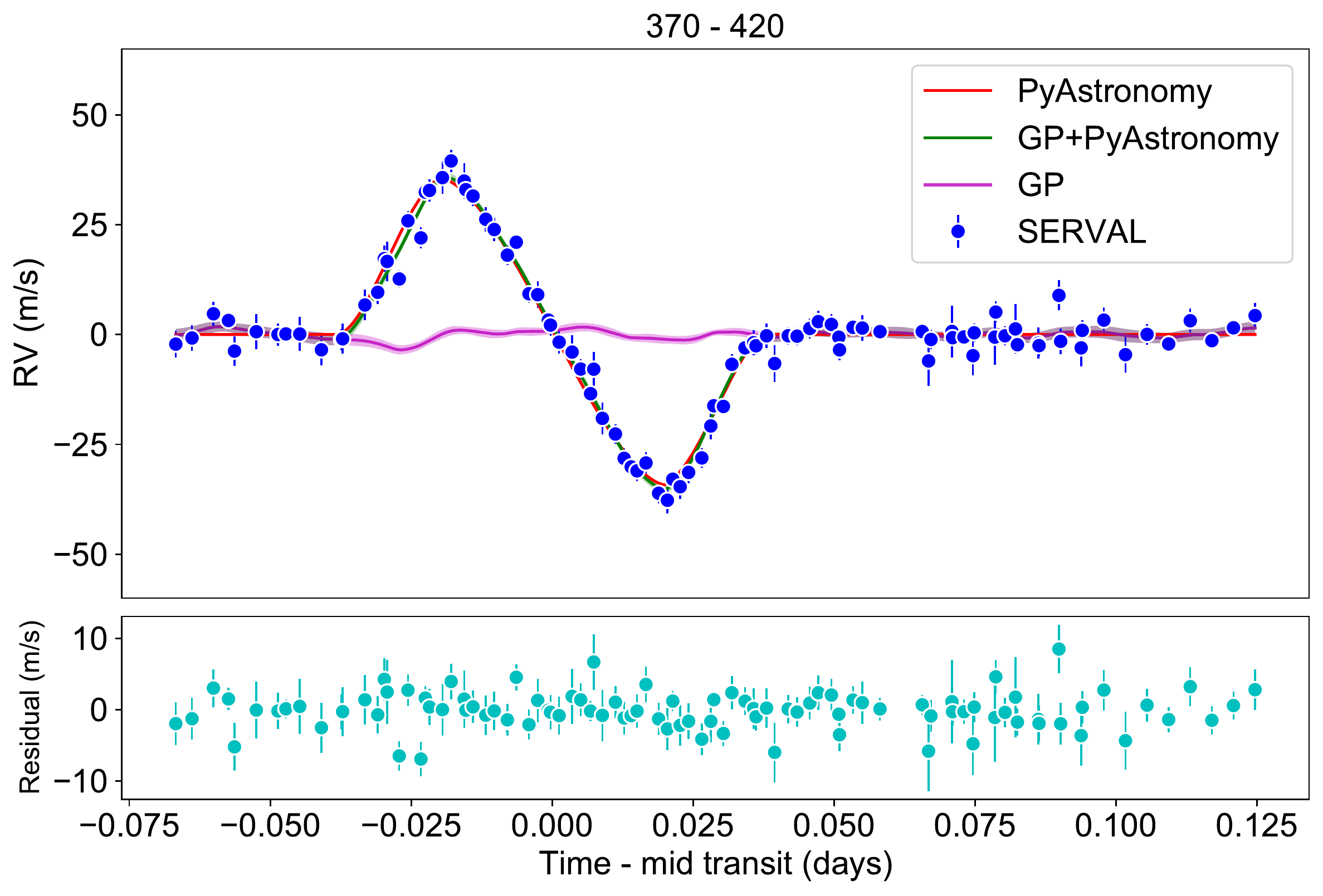}}
  \subfloat{\includegraphics[width=0.32\textwidth, height=4 cm]{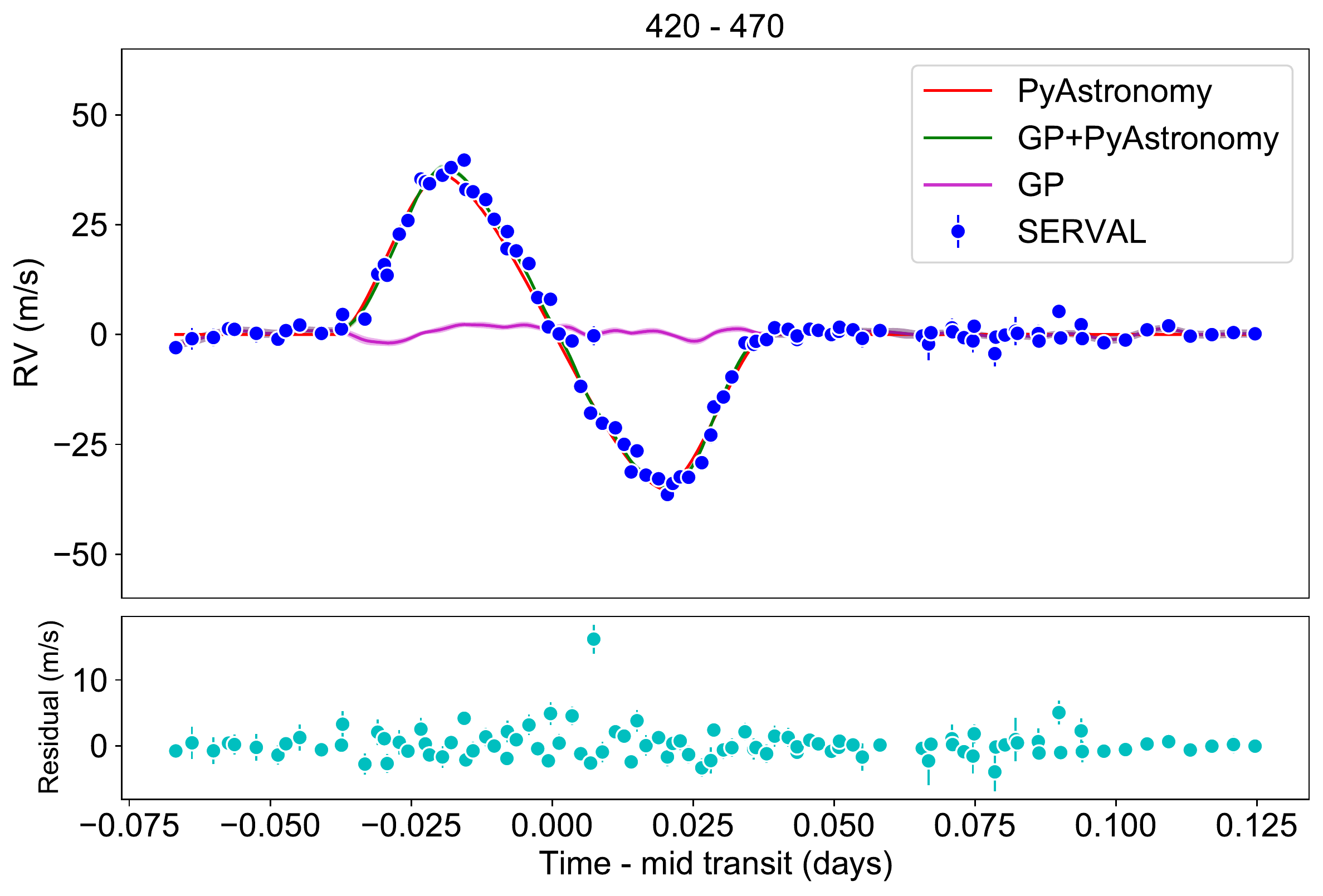}}
  \subfloat{\includegraphics[width=0.32\textwidth, height=4 cm]{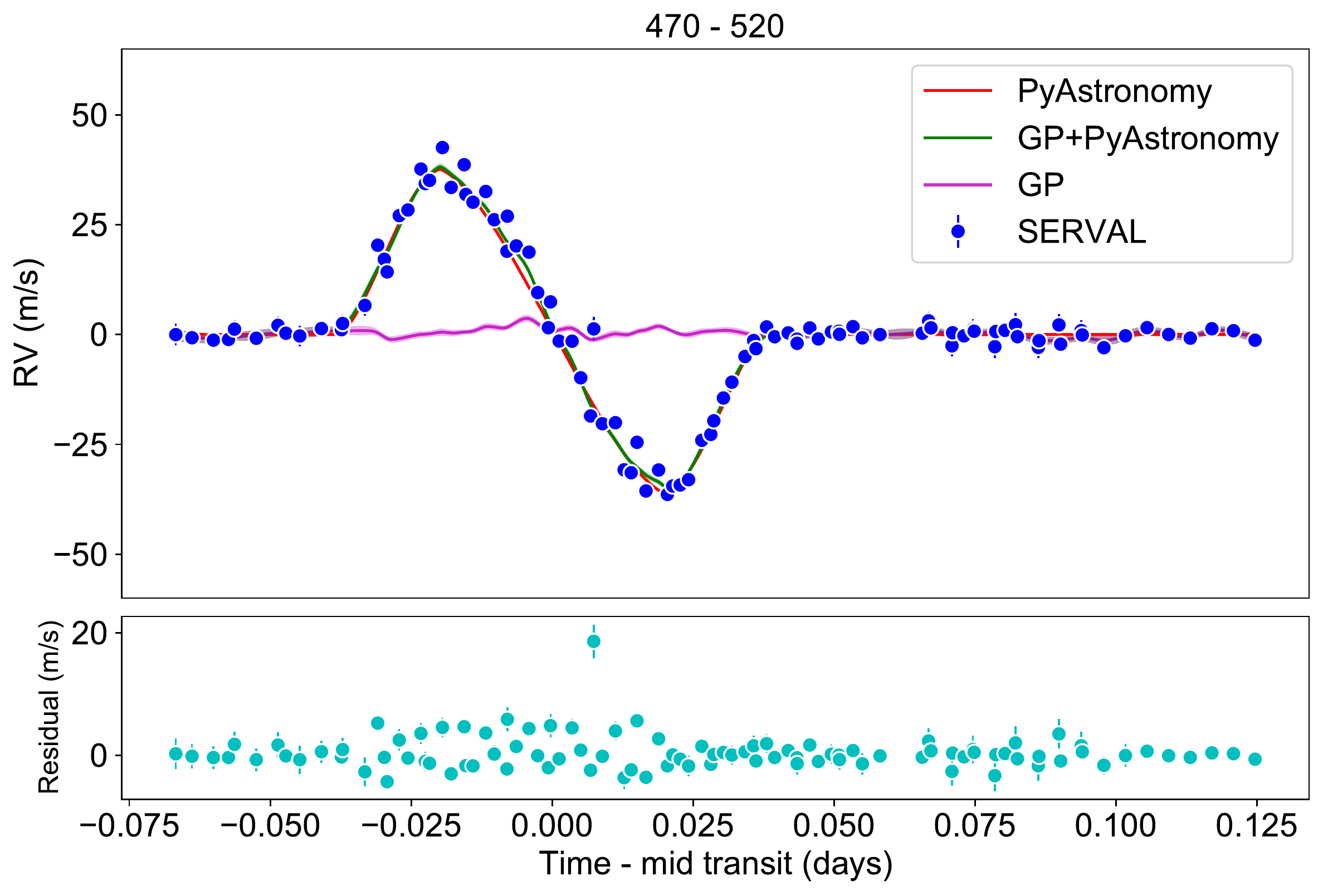}}\hfill
    \subfloat{\includegraphics[width=0.32\textwidth, height=4 cm]{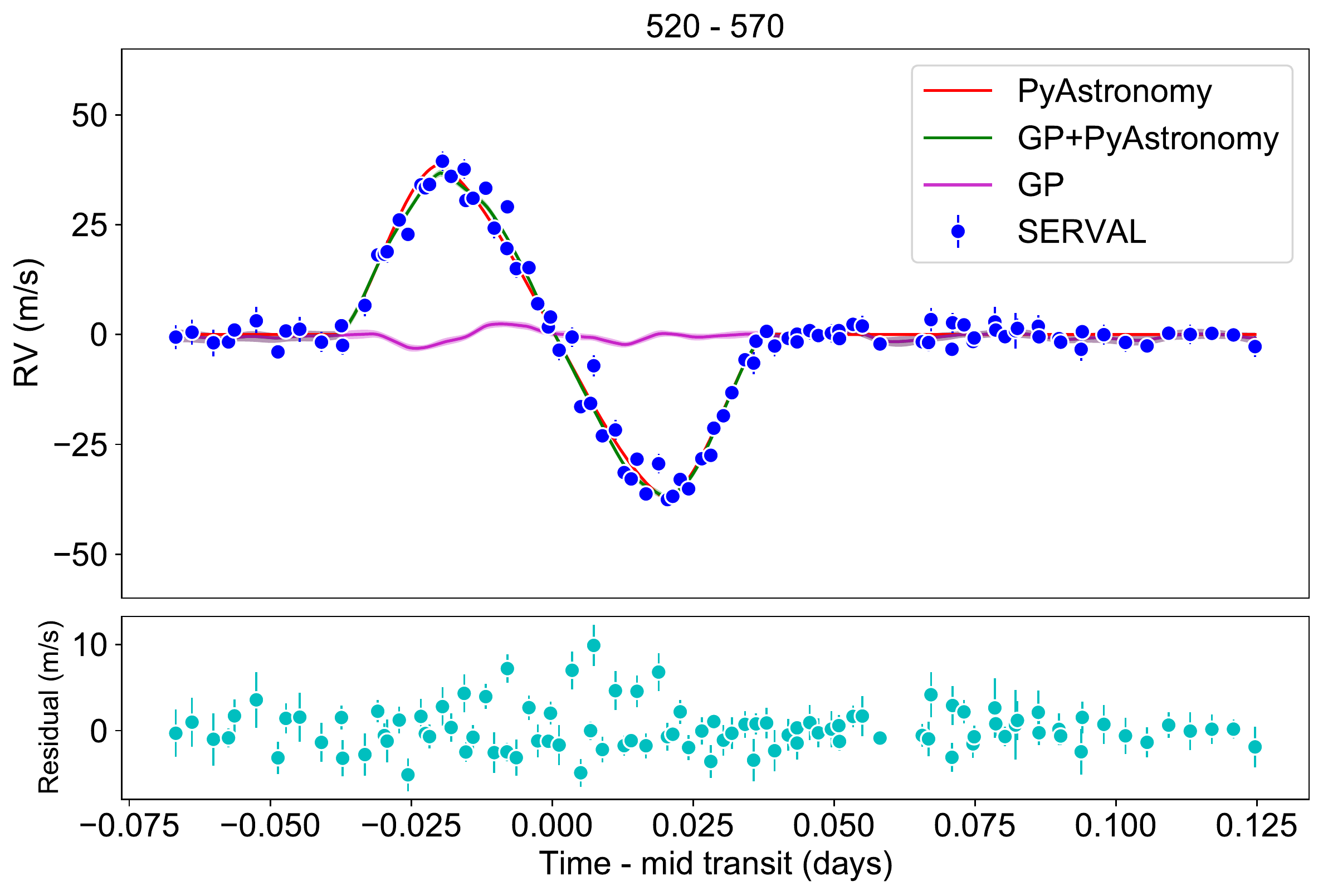}}
      \subfloat{\includegraphics[width=0.32\textwidth, height=4 cm]{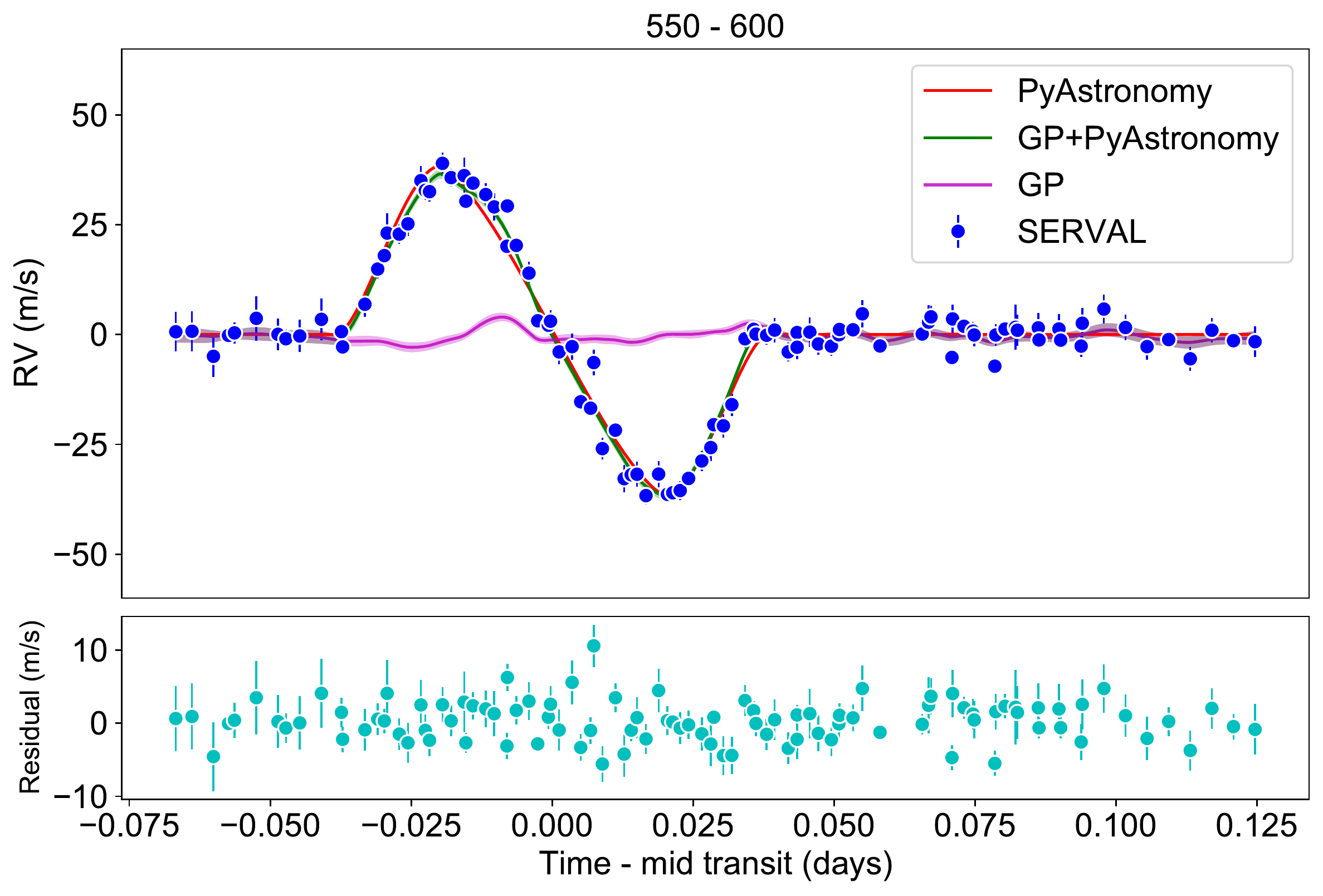}}
        \subfloat{\includegraphics[width=0.32\textwidth, height=4 cm]{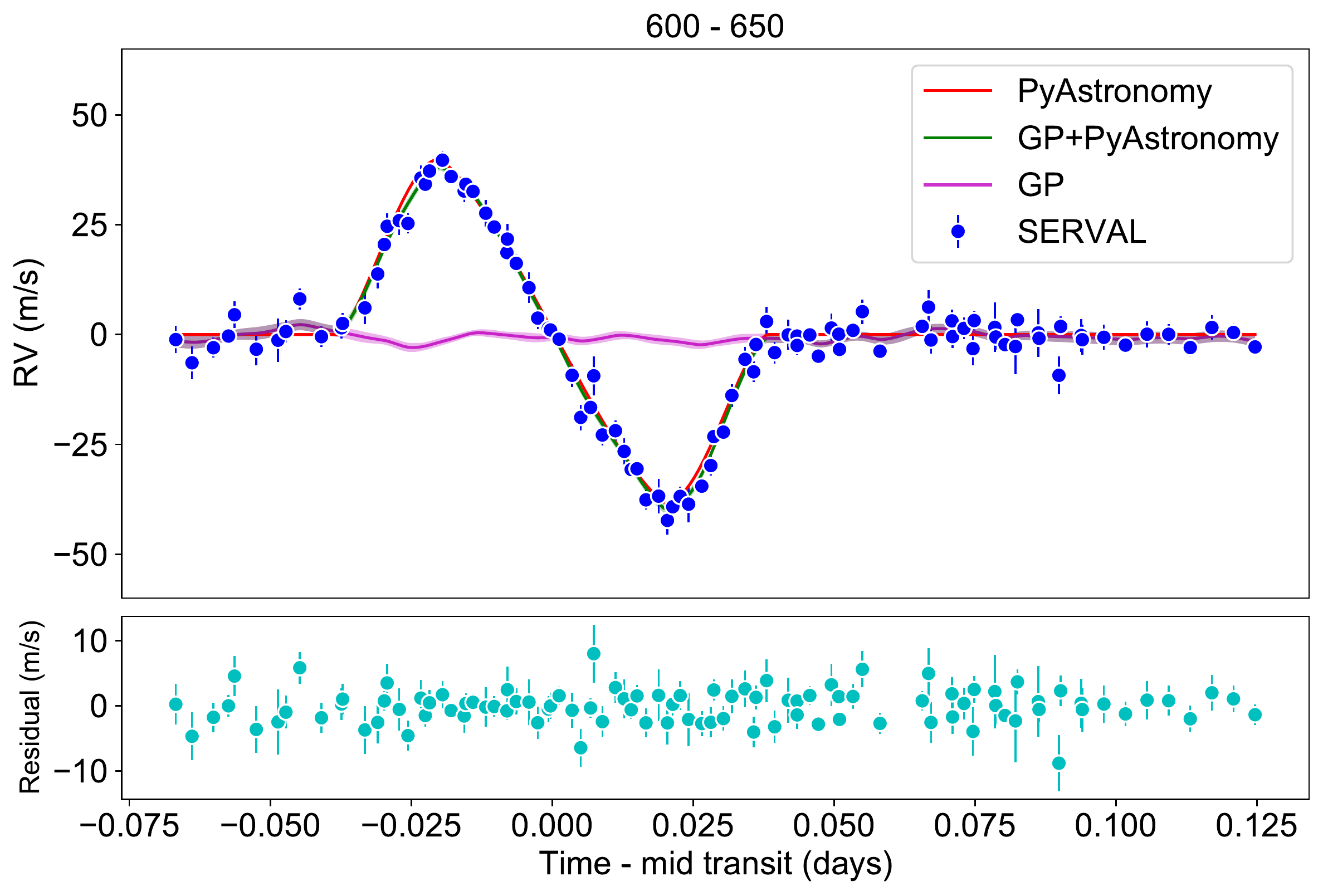}}\hfill
          \subfloat{\includegraphics[width=0.32\textwidth, height=4 cm]{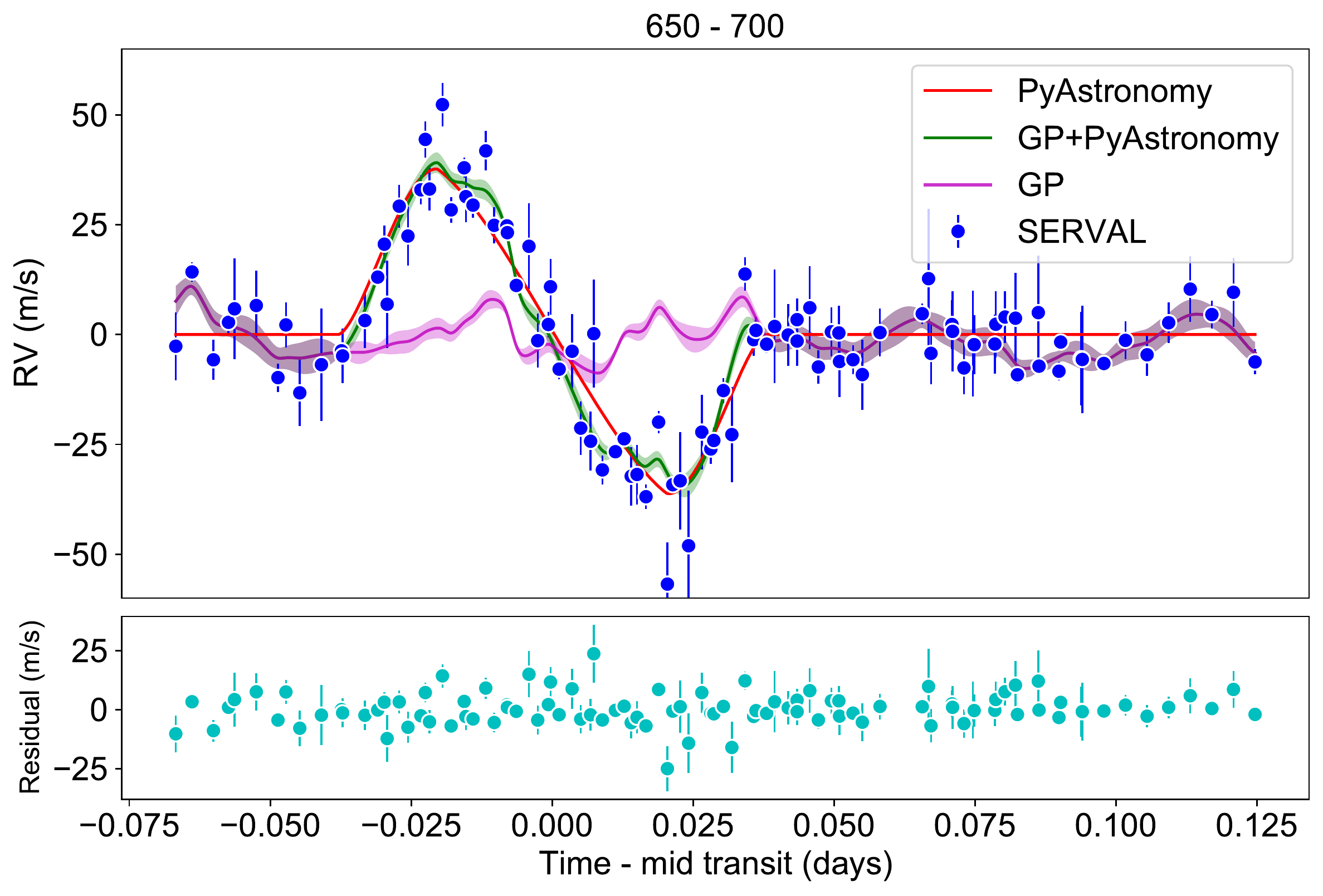}}
    \caption{RM curves derived with {\serval} in different wavelength bins using HARPS observations. The out-of-transit slope has been removed for each individual wavelength bin. The best-fit model to the RM using GP+\texttt{PyAstronomy} models is also shown. The different components of each best-fit model are plotted in different colors and marked in the legend. The title of each panel represents its corresponding wavelength range in nm.}%
      \label{fig:HARPS-BESTFIT}
\end{figure*}

\begin{figure*}
 \subfloat{\includegraphics[width=0.32\textwidth, height=4 cm]{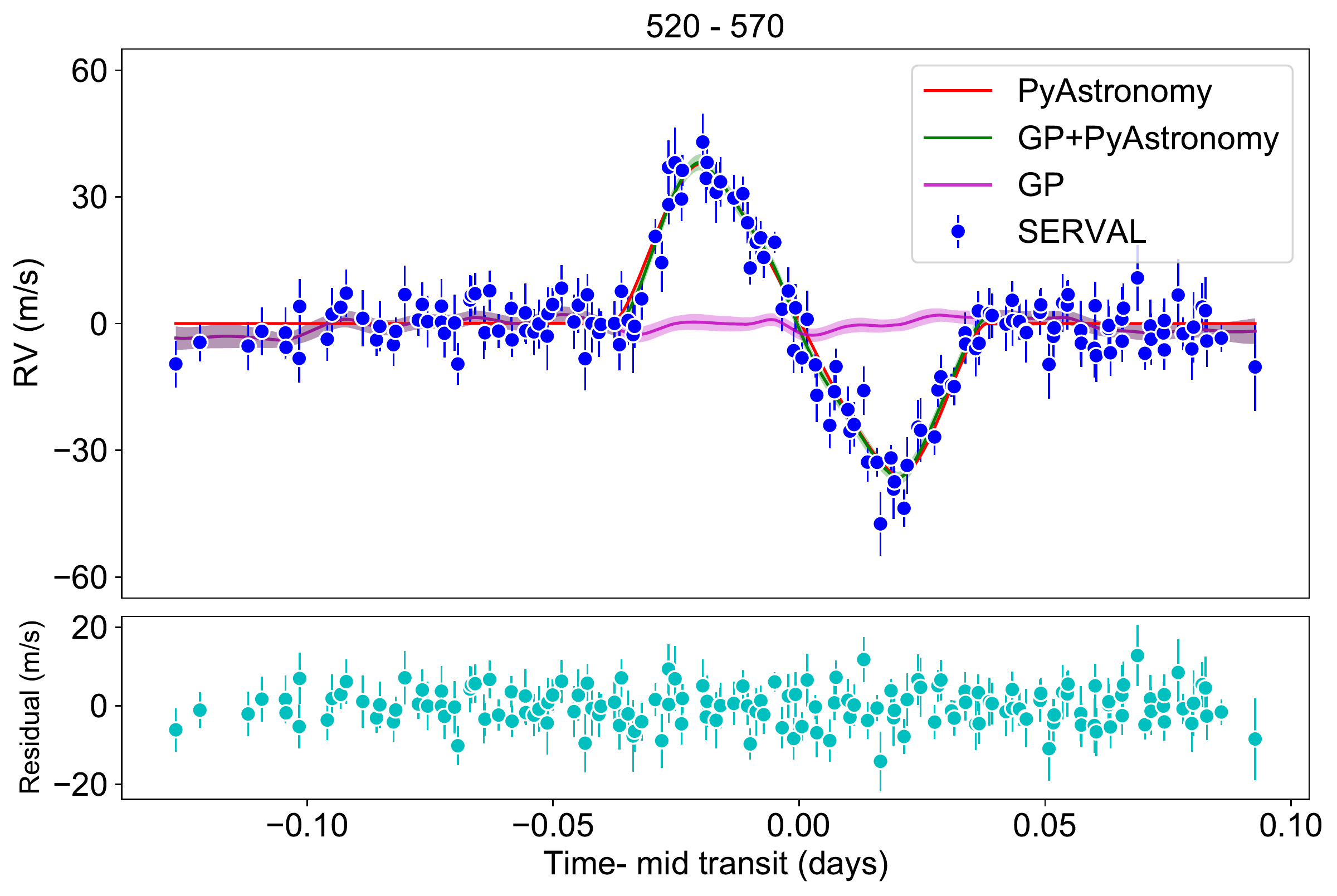}}
  \subfloat{\includegraphics[width=0.32\textwidth, height=4 cm]{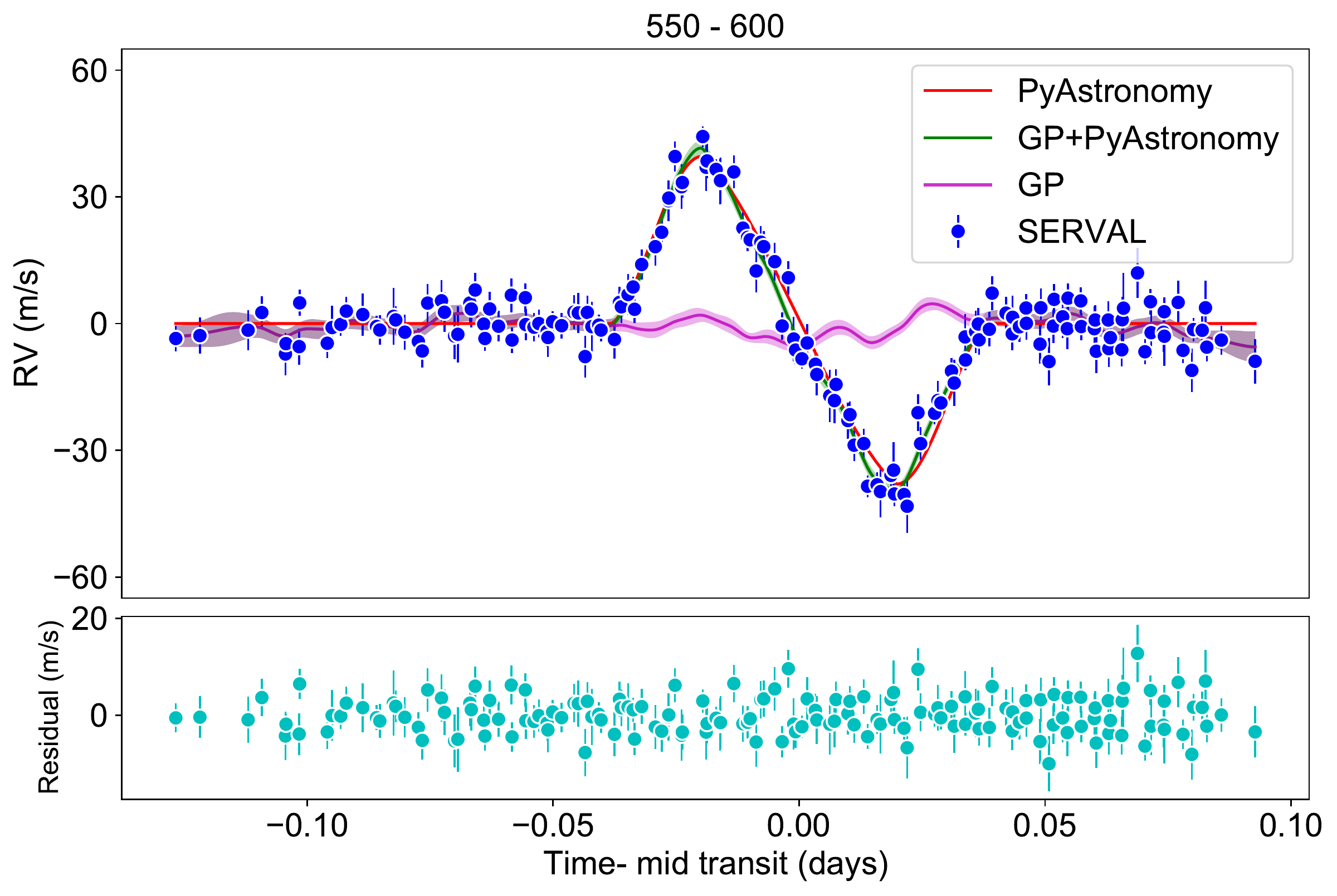}}
  \subfloat{\includegraphics[width=0.32\textwidth, height=4 cm]{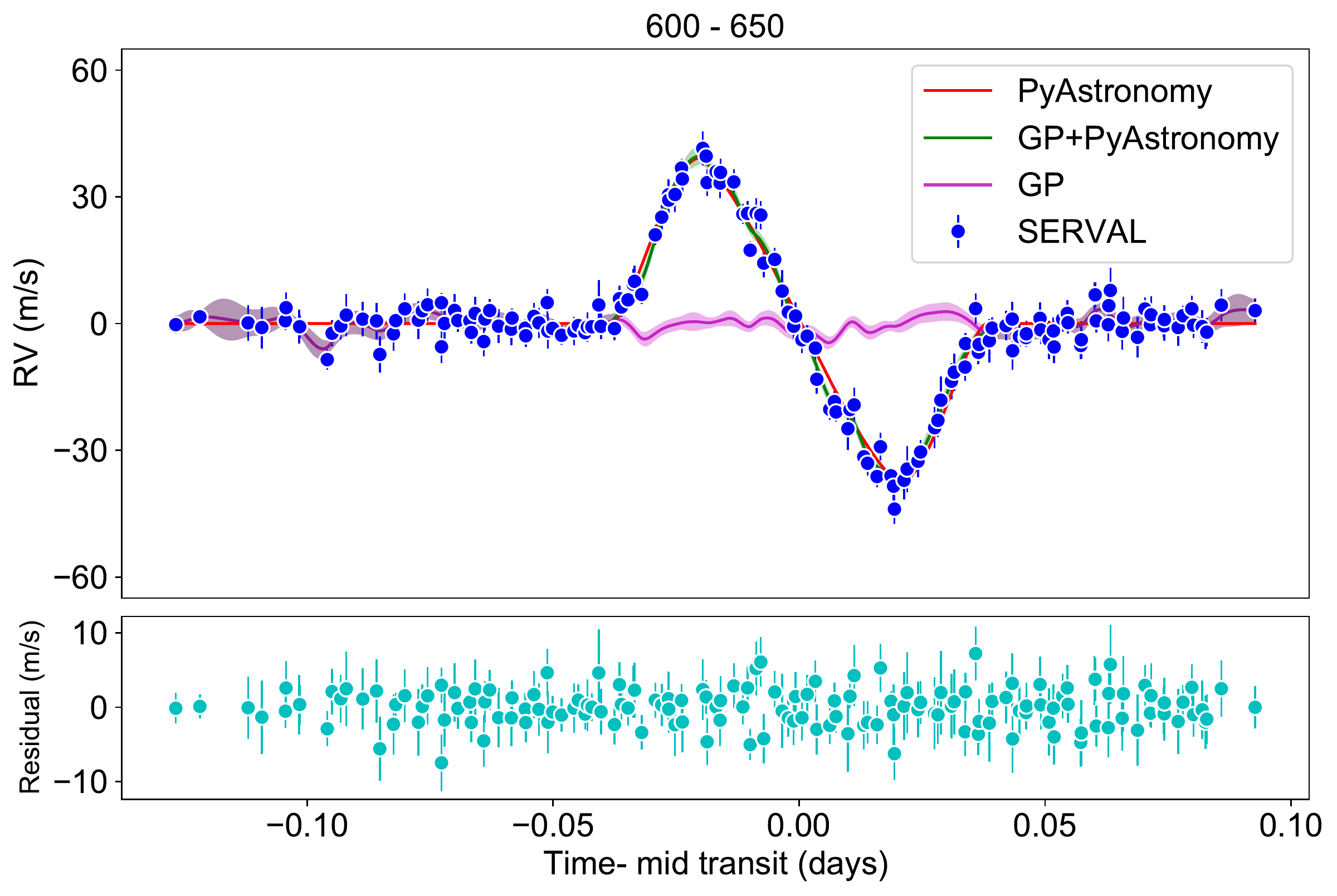}}\hfill
    \subfloat{\includegraphics[width=0.32\textwidth, height=4 cm]{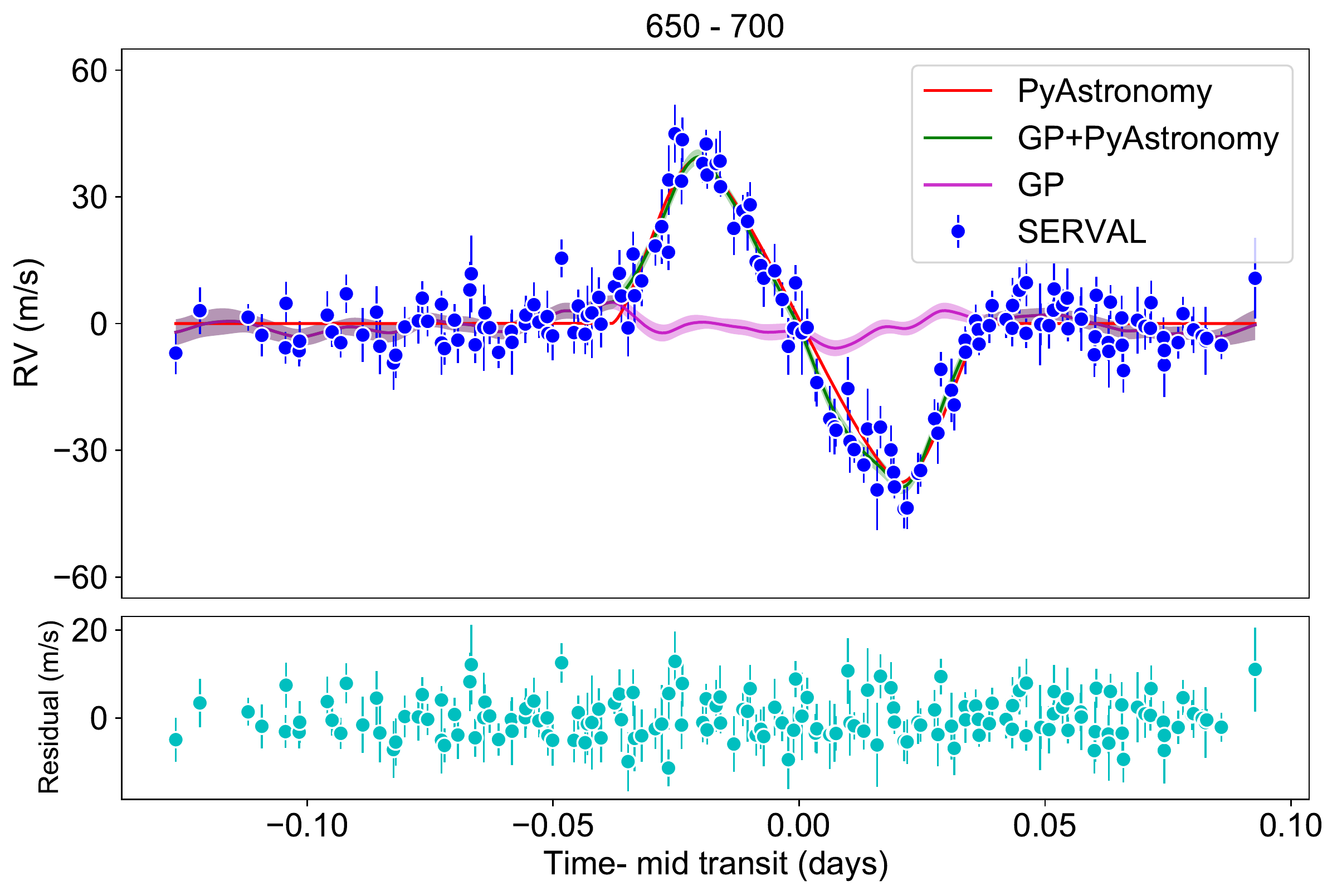}}
      \subfloat{\includegraphics[width=0.32\textwidth, height=4 cm]{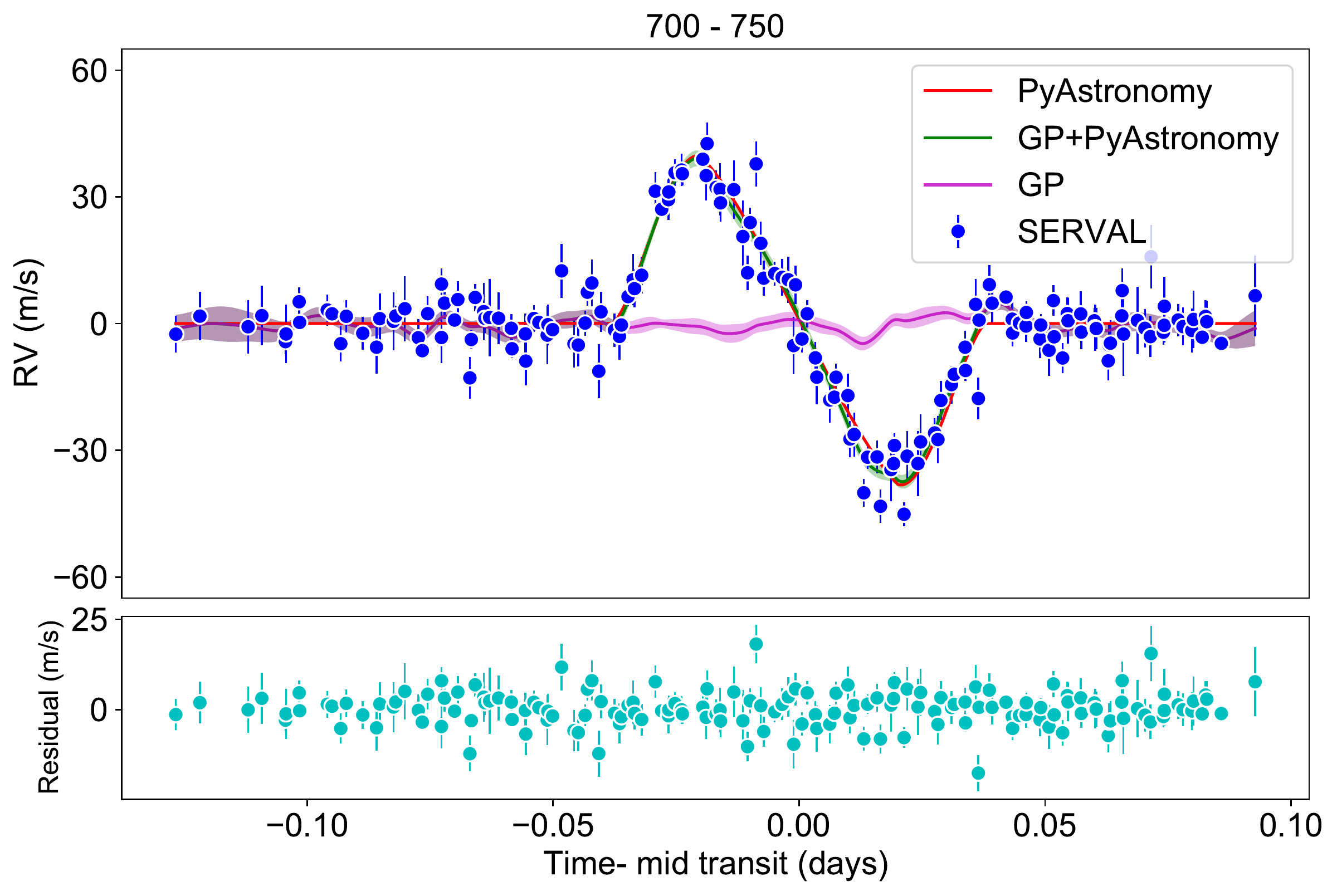}}
        \subfloat{\includegraphics[width=0.32\textwidth, height=4 cm]{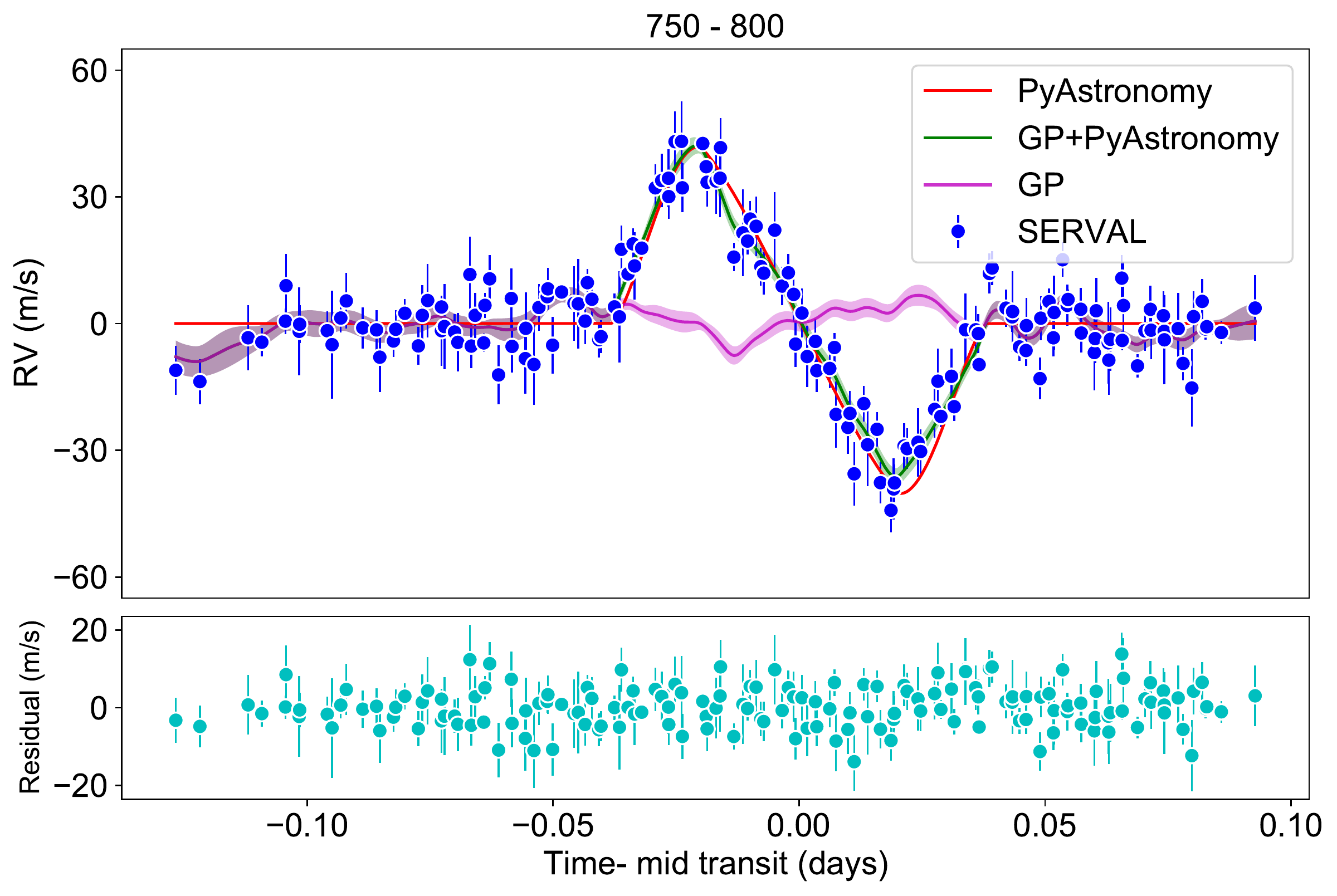}}\hfill
          \subfloat{\includegraphics[width=0.32\textwidth, height=4 cm]{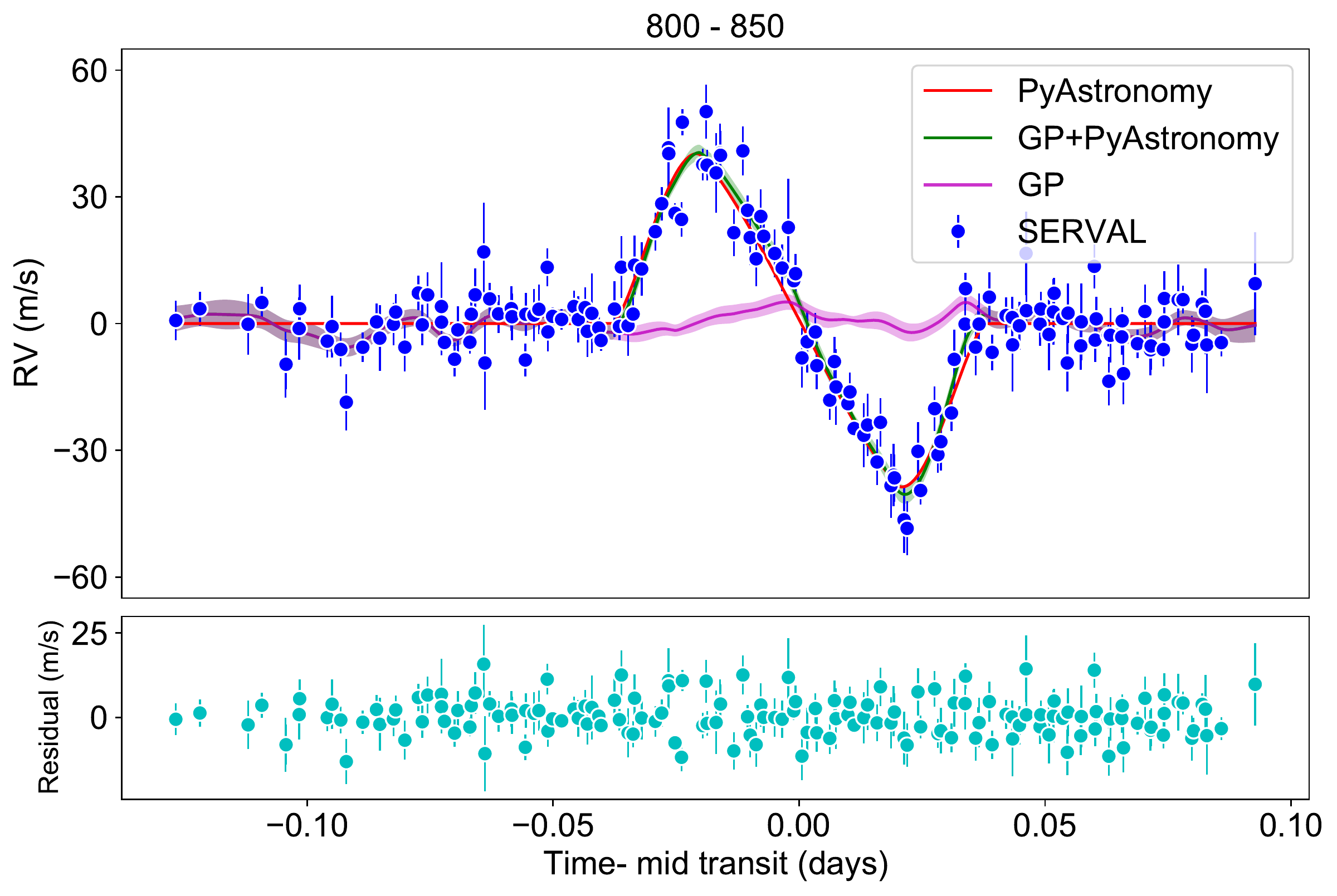}}
        \subfloat{\includegraphics[width=0.32\textwidth, height=4 cm]{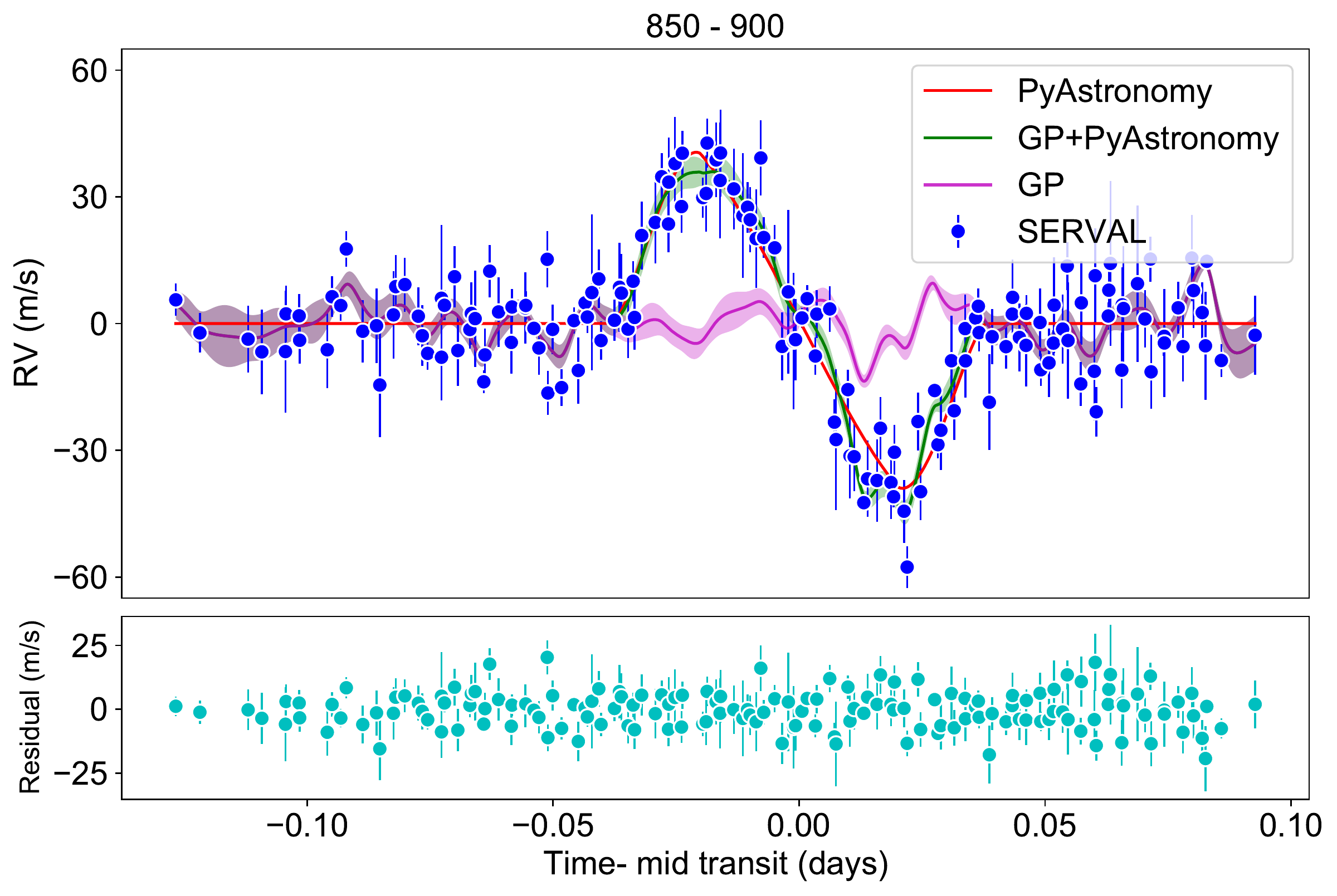}}
        \subfloat{\includegraphics[width=0.32\textwidth, height=4 cm]{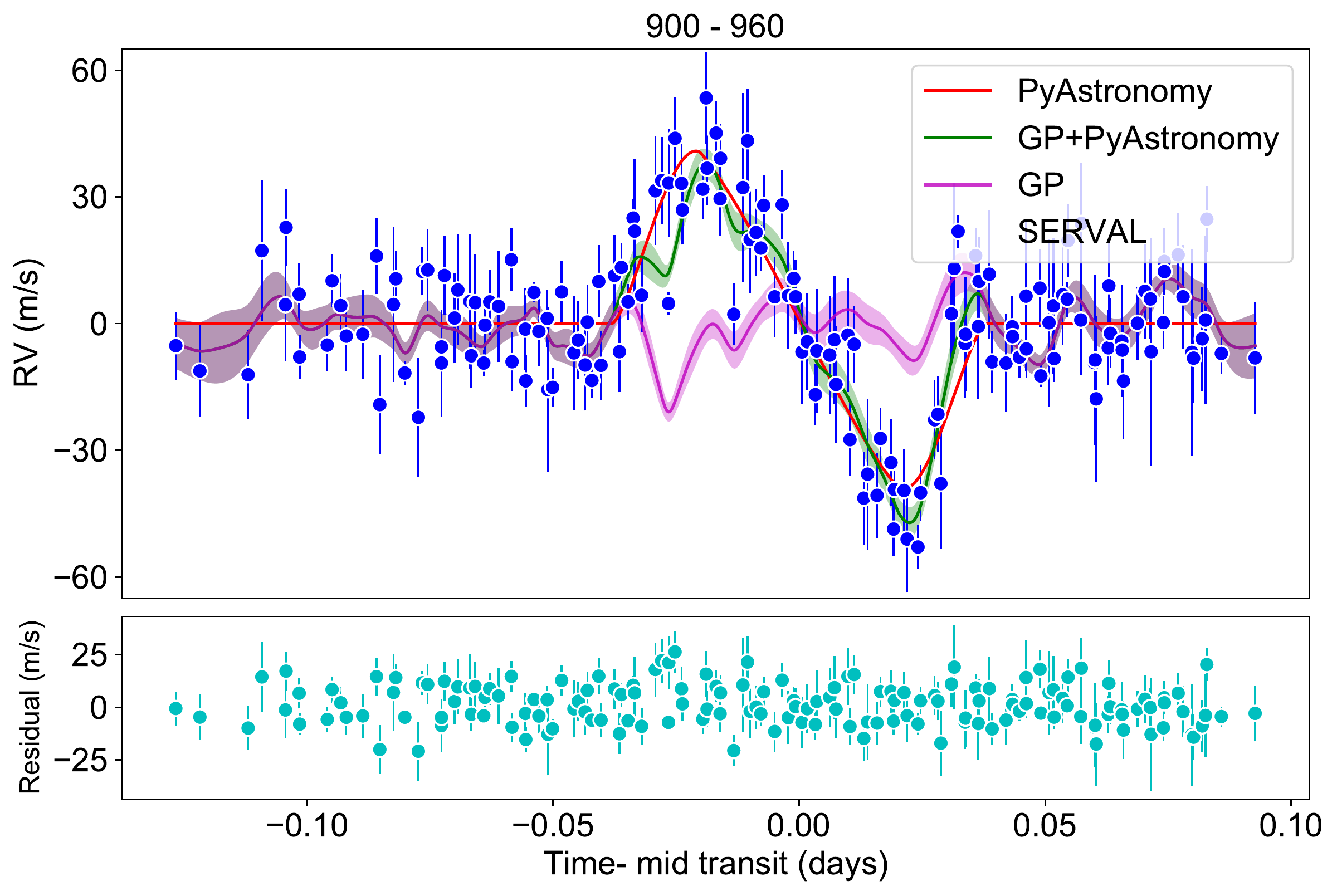}}
    \caption{Same as Fig.~\ref{fig:HARPS-BESTFIT}, but for CARMENES-VIS observations.}%
          \label{fig:CARMENES-VIS-BESTFIT}
\end{figure*}

\begin{figure*}
        \subfloat{\includegraphics[width=0.32\textwidth, height=4 cm]{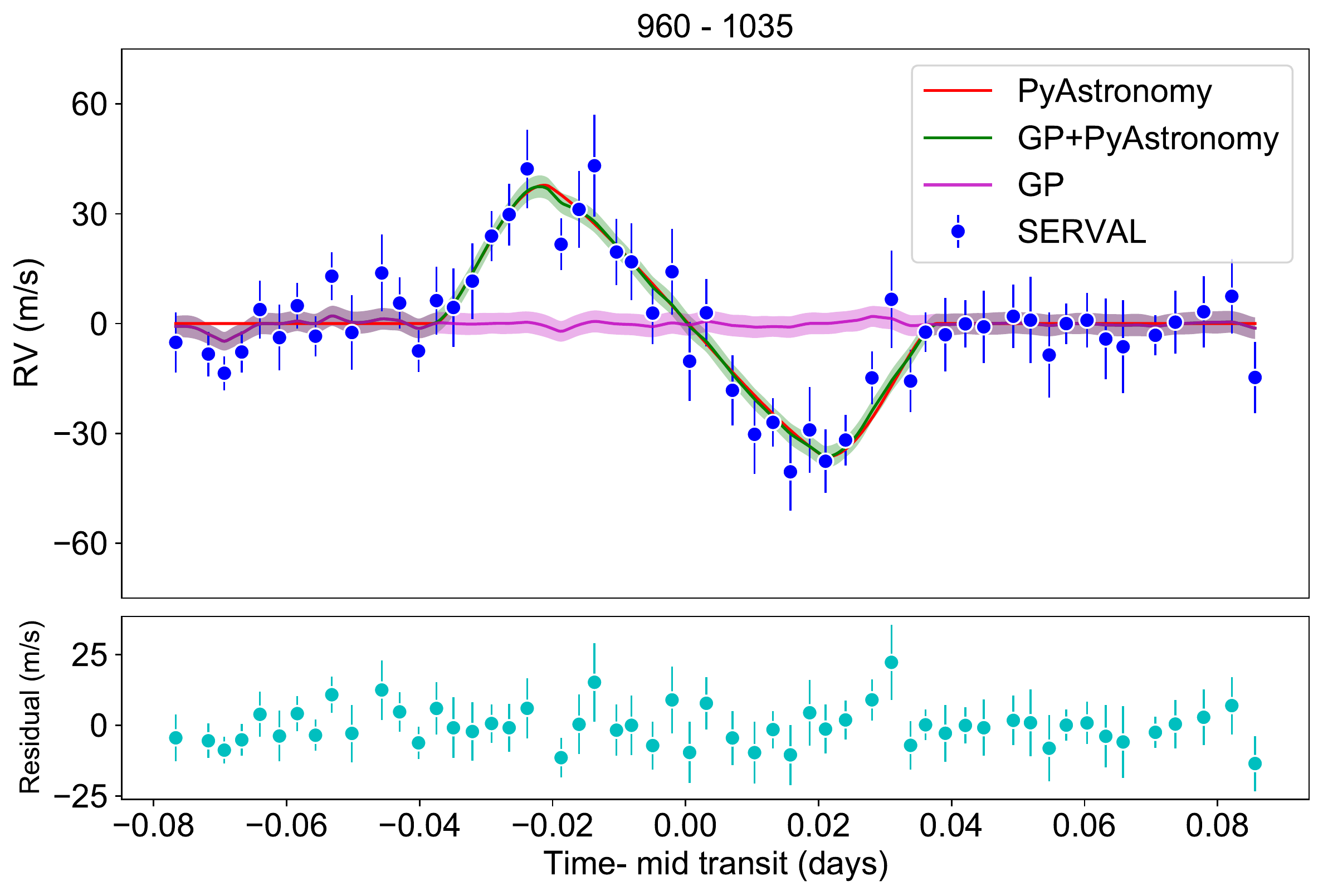}}
 \subfloat{\includegraphics[width=0.32\textwidth, height=4 cm]{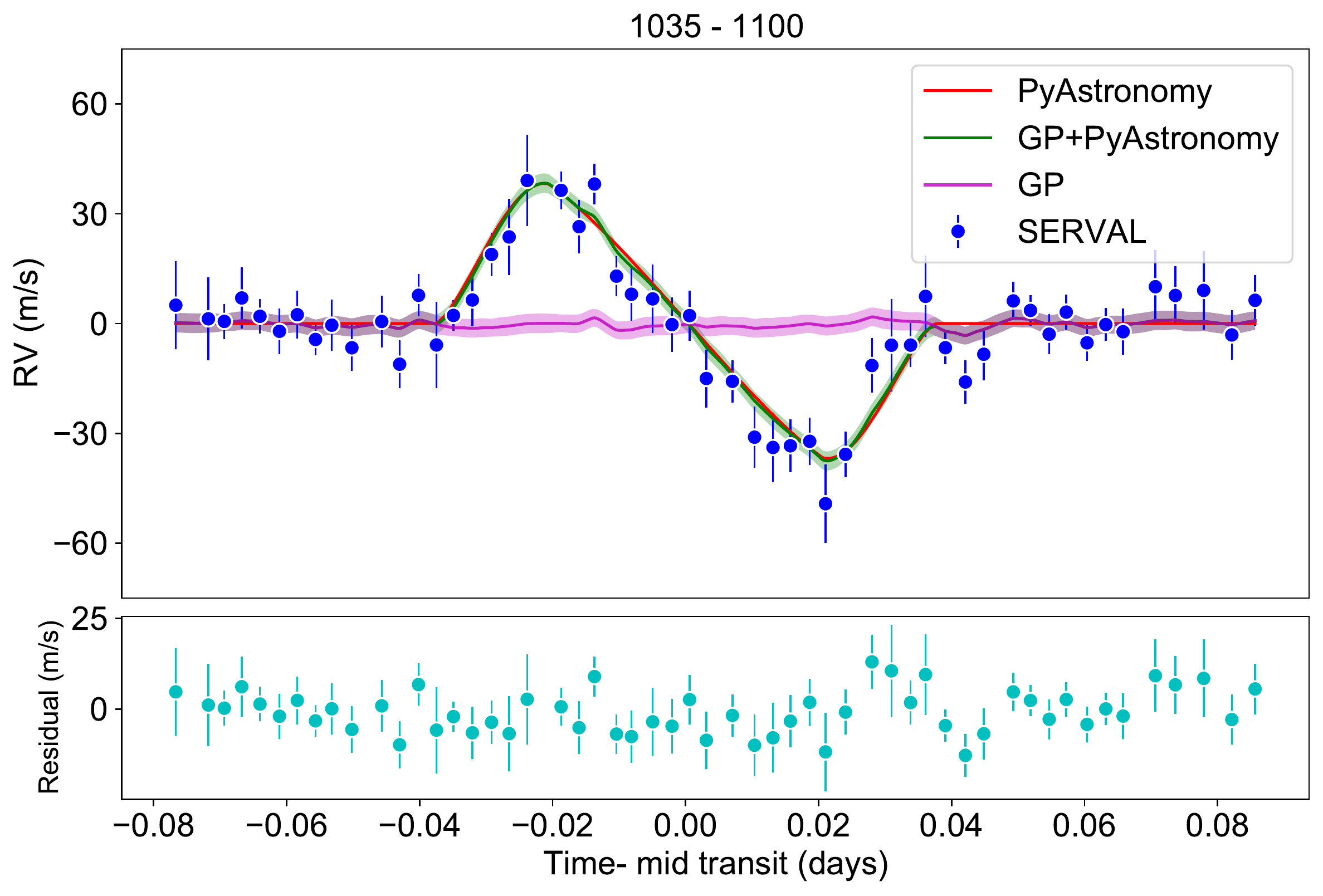}}
  \subfloat{\includegraphics[width=0.32\textwidth, height=4 cm]{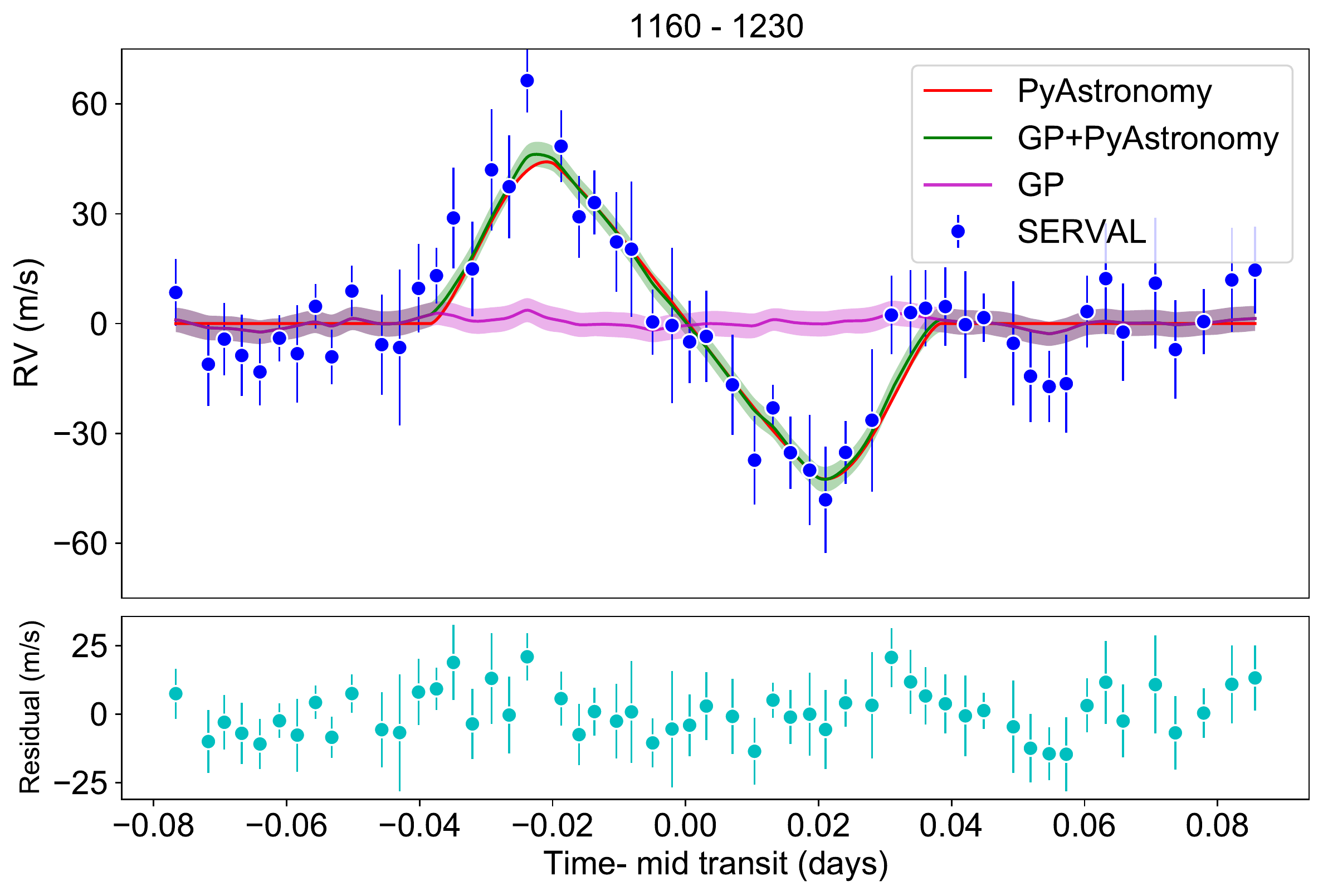}}\hfill
  \subfloat{\includegraphics[width=0.32\textwidth, height=4 cm]{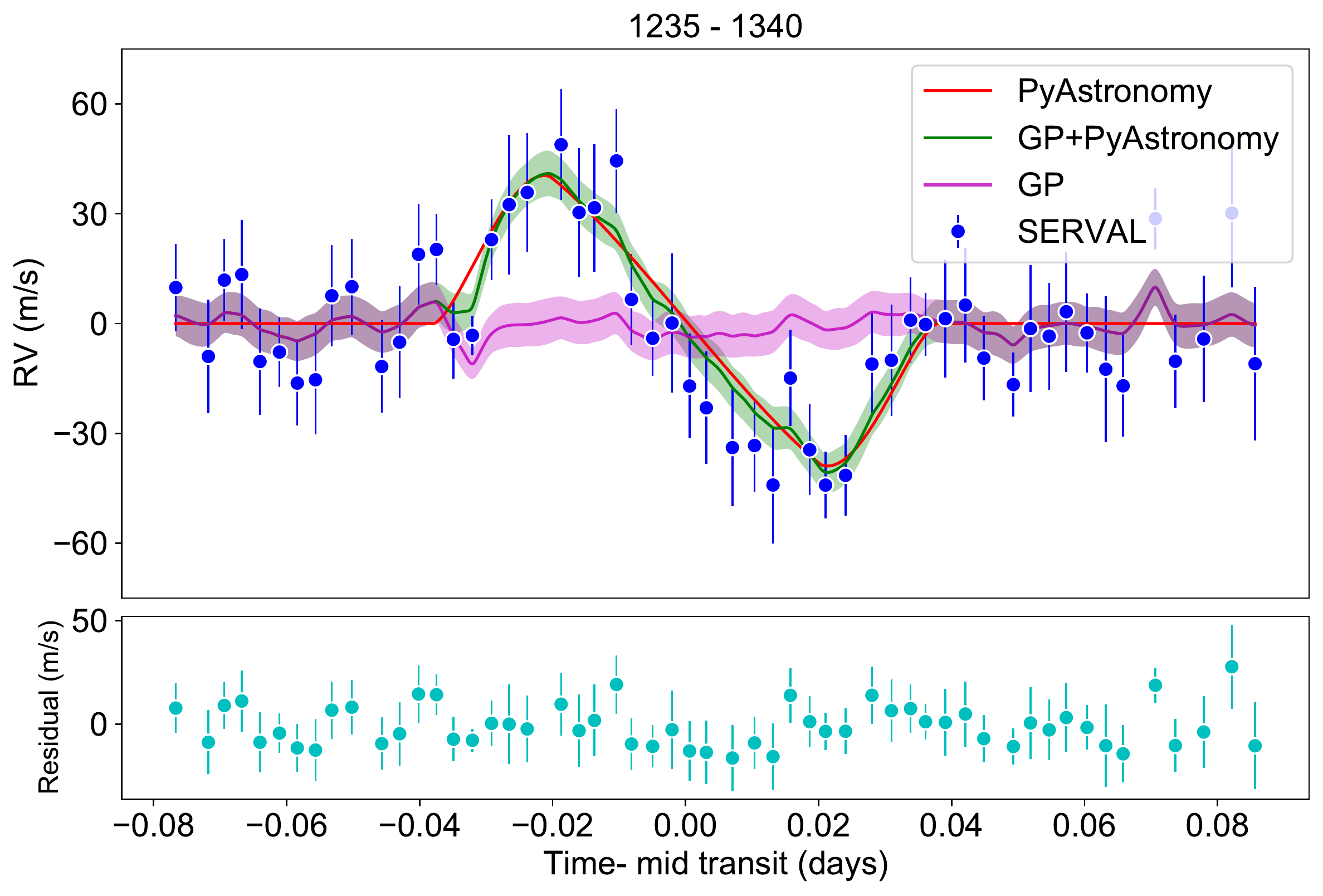}}
    \subfloat{\includegraphics[width=0.32\textwidth, height=4 cm]{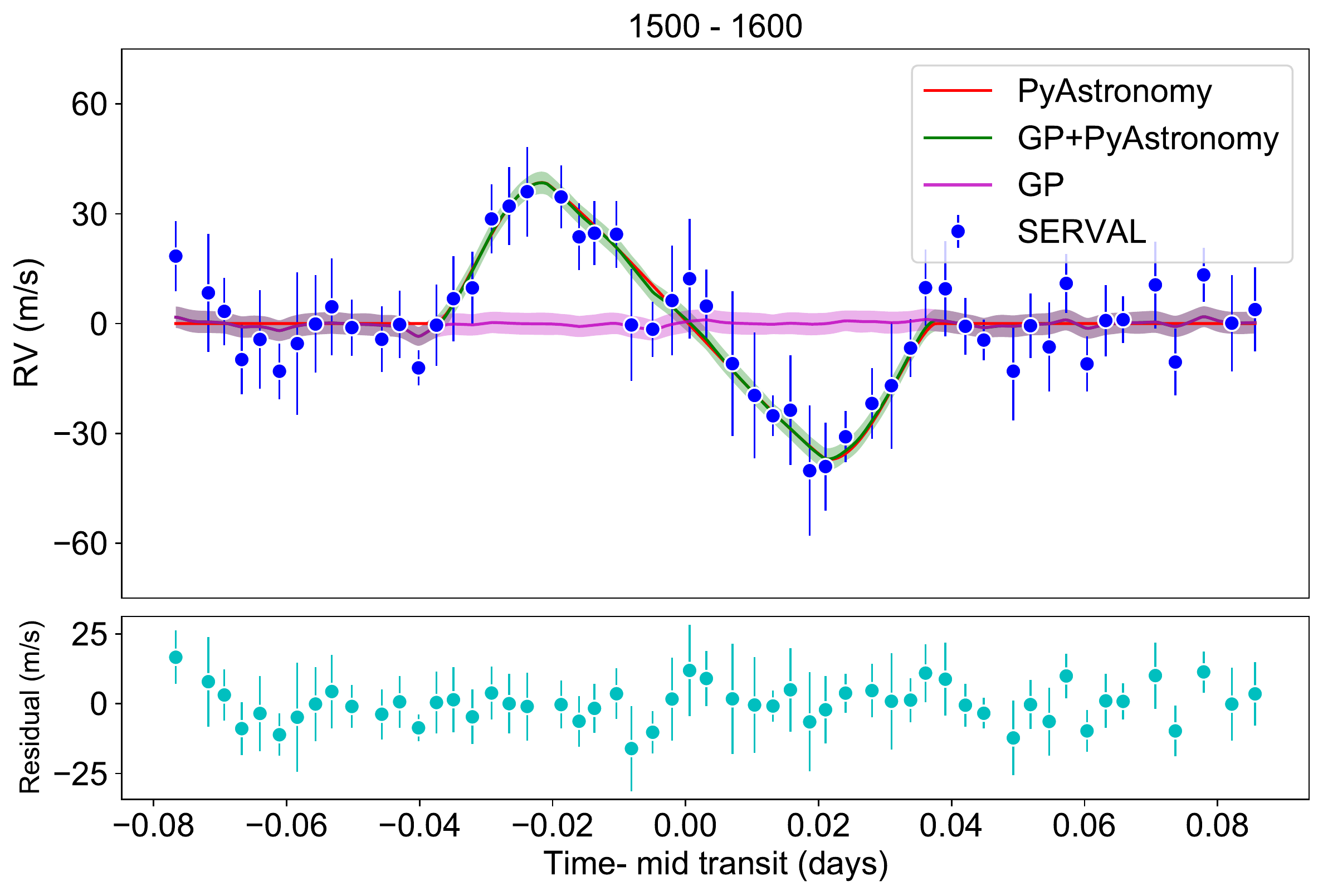}}
      \subfloat{\includegraphics[width=0.32\textwidth, height=4 cm]{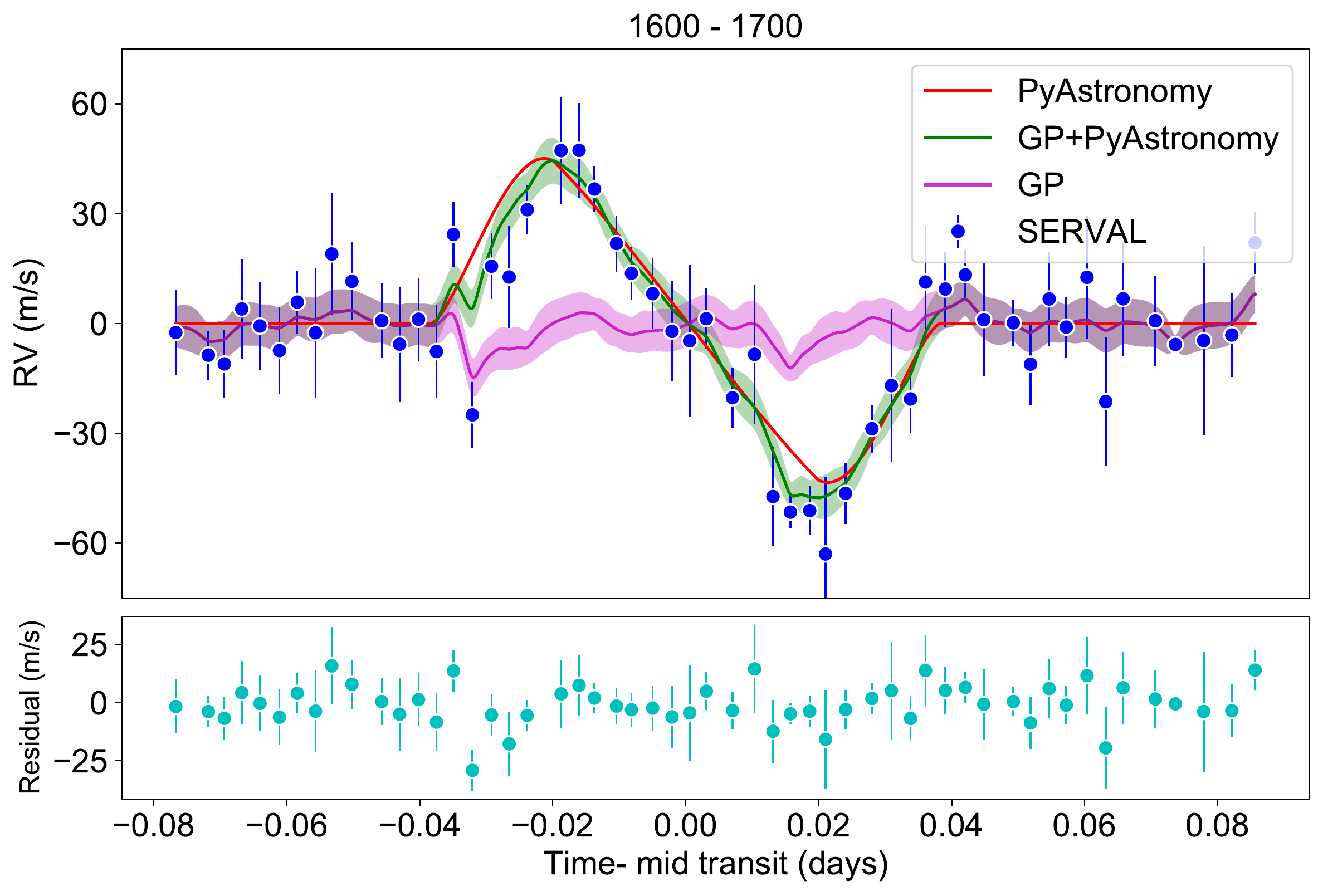}}

    \caption{Same as Fig.~\ref{fig:HARPS-BESTFIT}, but for CARMENES-NIR observations.}%
          \label{fig:CARMENES-NIR-BESTFIT}
\end{figure*}

\begin{figure*}
        \centering
        \includegraphics[width=0.7\linewidth, height=20cm]{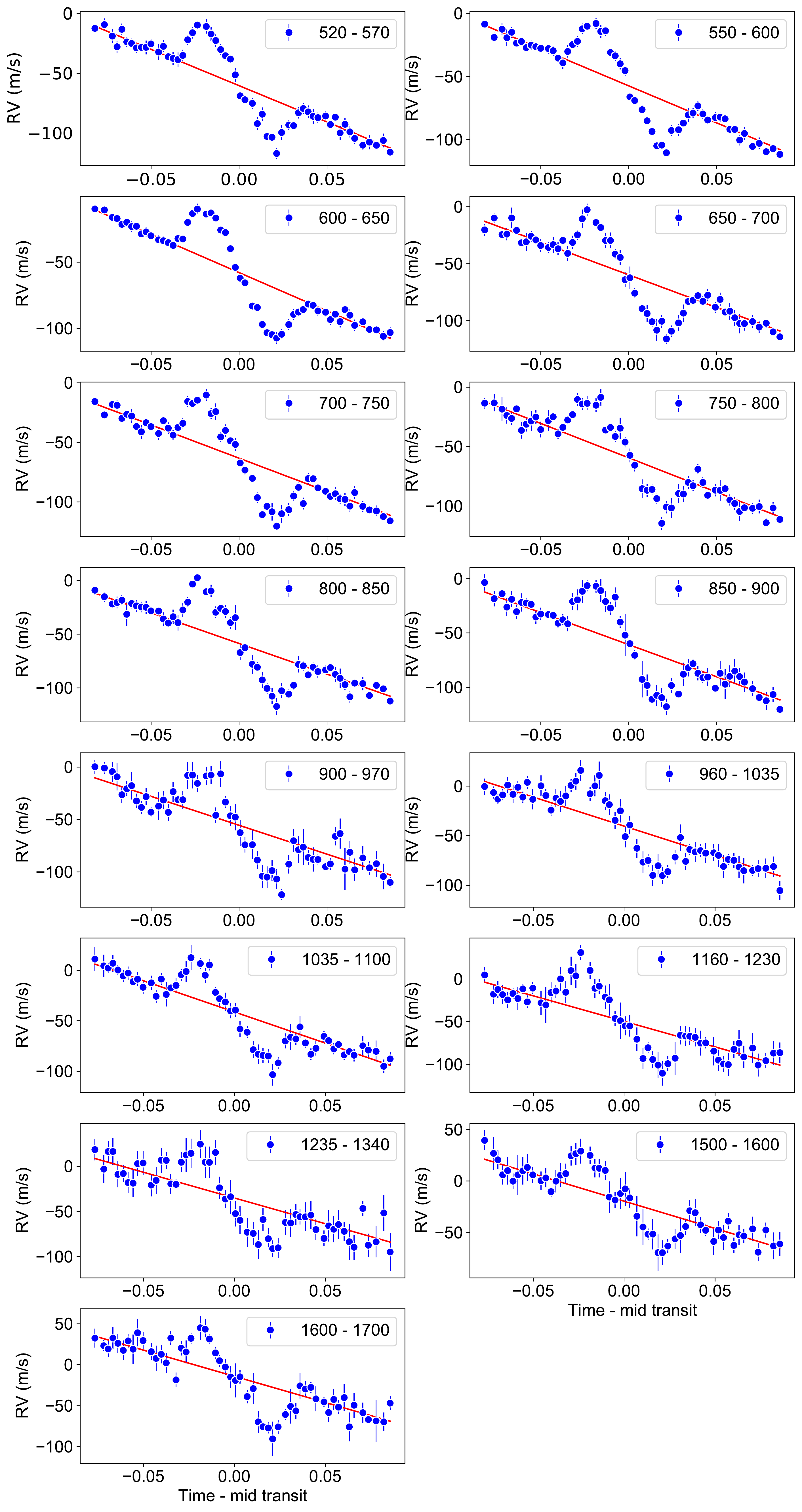}
        \caption{RM curves derived with {\serval} in different wavelength bins for CARMENES (VIS+NIR) observations during the transit of HD~189733b on 9 August 2019. The best-fit linear model to the out-of-transit observations is plotted as solid red lines. The legend of each panel represents its corresponding wavelength range in nm. }
        \label{fig:SLOPE}
\end{figure*}

\begin{figure*}
        \centering
        \includegraphics[width=1.\linewidth]{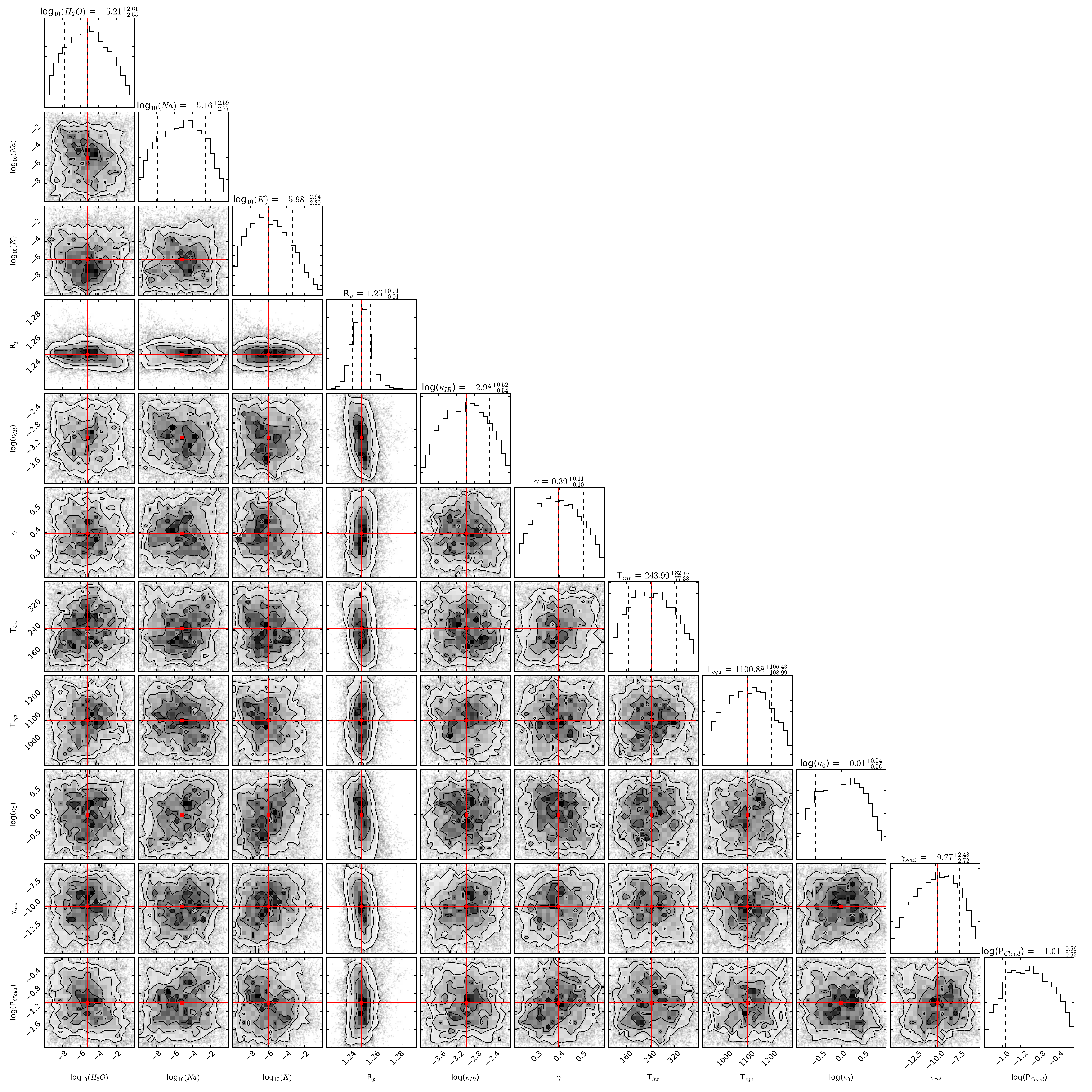}
        \caption{Retrieved posterior distributions by fitting atmospheric models to HARPS+CARMENES-VIS data.}
        \label{fig:retrieval_corner}
\end{figure*}

\end{document}